\documentclass[aps,twocolumn,longbibliography]{revtex4-1}
\usepackage{pdfpages}
\usepackage{amsmath,amssymb,braket,cases}
\usepackage{graphicx}
\usepackage{booktabs,multirow}
\usepackage{hyperref}
\usepackage{float}

\newcommand{\bk}{{\bf k}}
\newcommand{\br}{{\bf r}}
\newcommand{\mev}{{\mathrm{meV}}}
\newcommand{\im}{\mathrm{Im}\,}

\makeatletter
\AtBeginDocument{\let\LS@rot\@undefined}
\makeatother

\begin{document}
\title{Quantum oscillation from in-gap states and non-Hermitian Landau level problem}

\author{Huitao Shen}
\affiliation{Department of Physics, Massachusetts Institute of Technology, Cambridge, Massachusetts 02139, USA}
\author{Liang Fu}
\affiliation{Department of Physics, Massachusetts Institute of Technology, Cambridge, Massachusetts 02139, USA}

\begin{abstract}
Motivated by recent experiments on Kondo insulators, we theoretically study quantum oscillations from disorder-induced in-gap states in small-gap insulators. By solving a non-Hermitian Landau level problem that incorporates the imaginary part of electron's self-energy, we show that the oscillation period is determined by the Fermi surface area in the absence of the hybridization gap, and derive an analytical formula for the oscillation amplitude as a function of the indirect band gap, scattering rates, and temperature. Over a wide parameter range, we find that the effective mass is controlled by scattering rates, while the Dingle factor is controlled by the indirect band gap. We also show the important effect of scattering rates in reshaping the quasiparticle dispersion in connection with angle-resolved photoemission measurements on heavy fermion materials.
\end{abstract}

\maketitle

Quasiparticles in interacting/disordered systems generally have a finite lifetime due to the presence of electron-electron, electron-phonon or electron-impurity scattering. The decay of quasiparticles is formally described by the imaginary part of electron's self-energy. In small-gap systems, the decay of a quasiparticle can alter its energy-momentum dispersion significantly, for example, transform two-dimensional Dirac points into  ``bulk Fermi arcs'' \cite{Kozii2017, Papaj2018}. 

In this work, we study the effect of quasiparticle lifetime on the quantum oscillation in small-gap insulators. The oscillation of various physical quantities, such as the magnetic susceptibility and resistivity with respect to the magnetic field is usually regarded as a key characteristics of metals with a Fermi surface \cite{Shoenberg1984}. The period of the oscillation is determined by the Fermi surface area, and the amplitude decay with the temperature is determined by electron's effective mass. Intriguingly, recent experiments found quantum oscillations in heavy fermion materials $\mathrm{SmB}_6 $ \cite{Li2014,Tan2015} and $\mathrm{YbB}_{12} $ \cite{Li2017,Hsu2018}, which are Kondo insulators with a small energy gap. The physical origin of these quantum oscillations in insulators is hotly debated \cite{Knolle2015,Knolle2017a,Knolle2017,Zhang2016a,Baskaran2015,Erten2017,Grubinskas2017,Sodemann2017,Pal2017,Pal2017a,Pal2017b,Ram2017}.

Motivated by, but not limited to, these experiments on Kondo insulators, we theoretically study quantum oscillations from disorder-induced in-gap states in small-gap insulators. In a generic two-band model with a hybridization gap, disorder leads to finite quasiparticle lifetime 
and in-gap states. The spectrum and width of Landau levels in a magnetic field is calculated by solving a {\it non-Hermitian} Landau quantization problem that incorporates the imaginary part of electron's self-energy. The density of states inside the gap, which comes from the tails of broadened Landau levels, is found to exhibit oscillations periodic in $1/B$. The period is given by the Fermi surface in the absence of hybridization. An analytical formula is derived for the oscillation amplitude as a function of the indirect band gap, scattering rates and the temperature. For a wide range of parameters, the temperature dependence of the quantum oscillation is qualitatively similar to Lifshitz-Kosevich (LK) theory of normal metals \cite{Lifshitz1956,Shoenberg1984,Champel2001}. A key difference, however, is that the cyclotron mass in the LK factor is not the band mass, but depends on the scattering rate. Moreover, the oscillation amplitude at a fixed temperature, i.e., the Dingle factor, is controlled by the indirect band gap, when the scattering rate is small.

The peculiarity of quantum oscillation amplitude we found in small-gap insulators, where the scattering rate controls LK factor instead of Dingle factor, is quite the opposite to the case of normal metals, where the scattering rate controls Dingle factor instead of LK factor. This result is an important prediction of our theory. It contrasts clearly with quantum oscillations in clean insulators that lack in-gap states, where the amplitude of magnetization oscillation exhibits non-monotonous temperature dependence \cite{Knolle2015} or deviates from LK factors \cite{Kishigi2014}, and the oscillation of thermally averaged density of states exhibits thermal activation behavior and drops to zero, instead of saturates, in the zero temperature limit \cite{Kishigi2014,Zhang2016a,Knolle2017a}. 

\begin{figure}
\includegraphics[width=0.75\columnwidth]{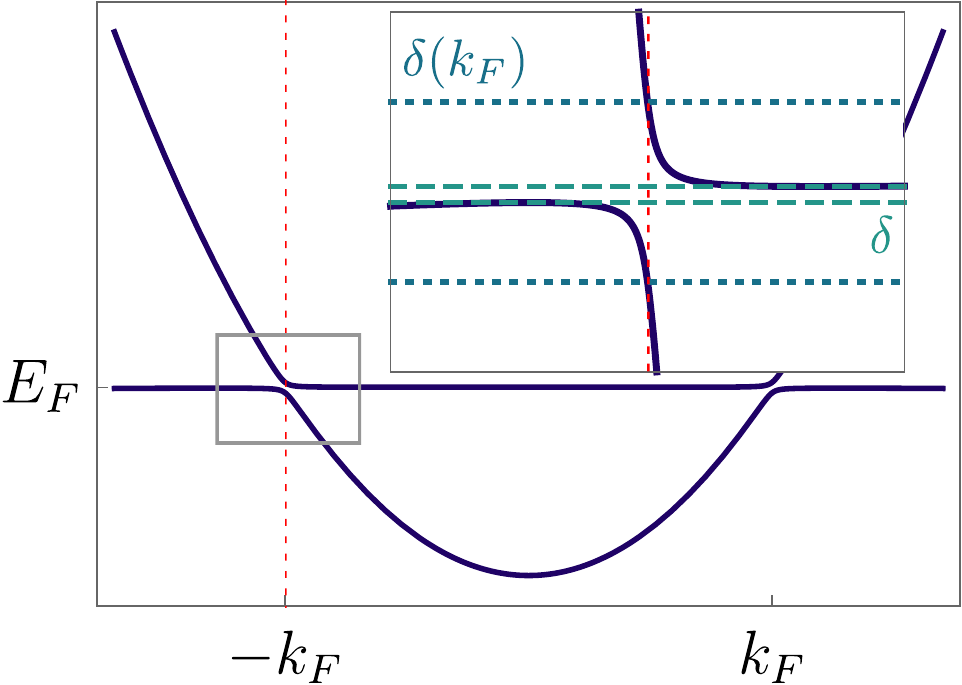}
\caption{Schematic band structure of a Kondo insulator. Inset: Zoom of the hybridization gap near $ k=k_F $. $ \delta(k_F) $ is the hybridization gap defined as the energy difference of the two bands at $ k=k_F $. $ \delta $ is the indirect band gap. }
\label{fig:gap}
\end{figure}

We start by considering a generic two-band model with a hybridization gap:
\begin{equation}
H_0(\bk)=\begin{pmatrix}
\epsilon_1(k) & \Delta(\bk) \\
\Delta(\bk) & -\epsilon_2(k)
\end{pmatrix}, \label{eq:h}
\end{equation}
with $ k\equiv | \bk |$. Diagonal terms $\epsilon_{1}(k)$ and $-\epsilon_2(k)$ describe the dispersion of an electron-type and a hole-type band respectively, and $\Delta(\bk)$ describes their hybridization.
This Hamiltonian in the inverted regime is widely used as a minimal model for the electronic structure of Kondo insulators at low temperatures \cite{Hewson1993}. In this context, the two bands come from $d$- and $f$-orbitals, and exhibit an avoided crossing on a circle or a sphere in $k$ space $k = k_F$, which is set by the condition $\epsilon_1(k_F) + \epsilon_2(k_F)=0$. Note that the hybridization gap $\delta(k_F)\equiv|\Delta(k_F)|$ is (much) larger than the indirect band gap $ \delta $, when the two bands are (highly) asymmetric, as shown in Fig.~\ref{fig:gap}.

Disorder introduces in-gap states in the above model. The disordered Hamiltonian we shall study is
$
H= \sum_\bk c^\dagger_\bk H_0(\bk) c_\bk + \int d\br U(\mathbf{r}) c^\dagger_\br \Lambda c_\br,
$
where $c^\dagger \equiv (d^\dagger, f^\dagger)$ is the electron creation operator for the two orbitals. $ U(\br) $ is the impurity potential, which is assumed to be characterized by $ \braket{U(\br)}=0 $ and $ \braket{U(\br)U(\br')}=n_{\rm imp}U_0^2\delta(\br-\br') $ under disorder average. $ n_{\rm imp} $ is the impurity density. The electron-impurity scattering is allowed to be orbital dependent, so the scattering vertex takes the form $\Lambda= \alpha I + \beta \sigma_z$.
In heavy fermion systems, $f$-orbitals are tightly bound to the nucleus and scatter much less with impurities than $d$-orbitals do.

Using self-consistent Born and T-matrix approximation, we compute disorder-induced electron self-energy operator $\Sigma(\omega)$, which is a $ 2\times 2 $ matrix. In systems with $ p $-wave hybridization, the self-energy is guaranteed to be diagonal \cite{SM}. The real part of $\Sigma(\omega)$ renormalizes the chemical potential and the inverted gap at $k=0$, and for convenience, will be absorbed into $ H_0 $ in the following.
The imaginary part of the self-energy becomes a nonzero diagonal matrix when the disorder strength $ n_{\rm imp} U_0 $ exceeds a critical value on the order of hybridization gap $\delta(k_F)$. At low energy $|\omega|\lesssim \delta(k_F)$, $\im\Sigma(\omega)$ is weakly dependent on $\omega$, hence can be approximated by
\begin{equation}
\Sigma(\omega) \simeq \begin{pmatrix}
-i\Gamma_1 & 0 \\ 0 & -i\Gamma_2
\end{pmatrix} \equiv \frac{- i}{2} (\Gamma I + \gamma \sigma_z).
\label{eq:se}
\end{equation}
$ \Gamma_1, \Gamma_2>0 $ are the inverse lifetimes of quasiparticles on the $d$- and $f$-band respectively, and we have defined $ \Gamma\equiv \Gamma_1+\Gamma_2$ and $ \gamma\equiv \Gamma_1-\Gamma_2 $. Generally, $ \Gamma_1\neq \Gamma_2 $ or $\gamma \neq 0$, as the two bands have different masses and disorder potentials.

The imaginary part of electron's self-energy modifies and broadens the quasiparticle dispersion, and creates in-gap states. To see this, we compute the spectral function
$
A({\bk}, \omega) = -2{\rm Im} \left[ 1/(w- H_0(\bk) -\Sigma) \right]. 
$
For a given $\bk$, $A({\bk}, \omega)$ is a sum of Lorentzians associated with the poles of the Green's function ${\cal E}_\pm (\bk)$, which are complex eigenvalues of the non-Hermitian quasiparticle Hamiltonian $H(\bk) \equiv H_0(\bk) + \Sigma$. For our two-band model and self-energy defined by Eq.~\eqref{eq:h} and \eqref{eq:se}, the two eigenvalues ${\cal E}_\pm (\bk)$ are \cite{Kozii2017}
\begin{eqnarray}
{\cal E}_{\pm}(\bk)&=&\frac{1}{2}\Big(\epsilon_1(k) - \epsilon_2(k) -i\Gamma \nonumber \\
&\pm& \sqrt{(\epsilon_1(k) + \epsilon_2(k) - i\gamma)^2+ \Delta^2 (\bk) }\Big) \label{eq:dispersion}
\end{eqnarray}
The real part of ${\cal E}_\pm(\bk)$, denoted as $\epsilon_\pm(\bk)$, is the dispersion of quasiparticle conduction and valence band, while its imaginary part
determines the width of the broadened spectral function.  
In the special case of a single scattering rate $\Gamma_1=\Gamma_2$ or $\gamma=0$,  the imaginary part is a constant so that the original band dispersion of $H_0(\bk)$ is broadened uniformly.

\begin{figure}
\includegraphics[width=\columnwidth]{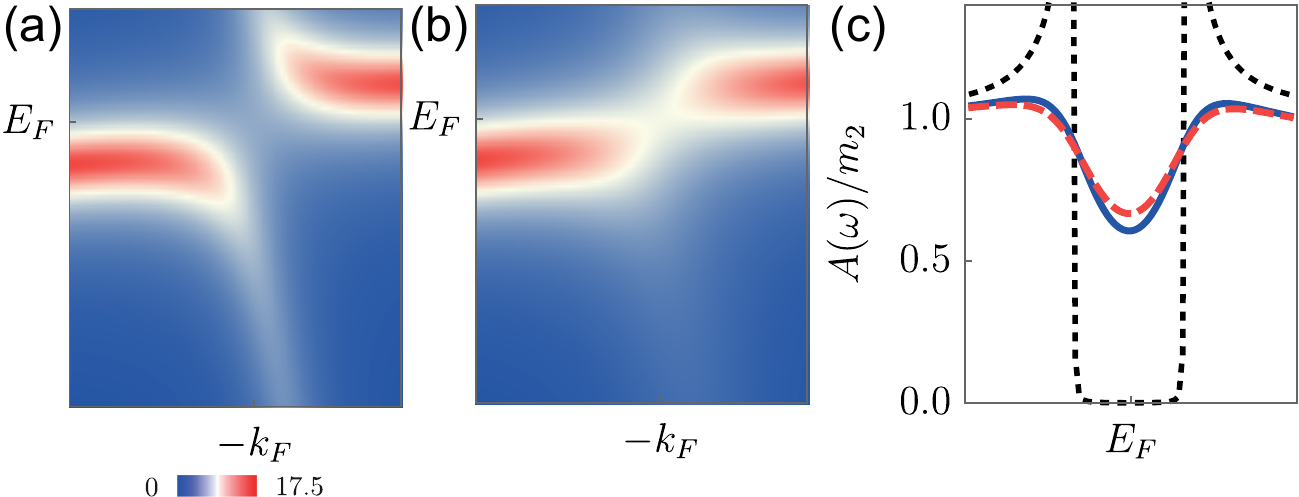}
\caption{(a) Spectral function $ A(k,\omega) $ for model $\epsilon_{i}=k^2/(2m_{i})- \mu_{i}$ ($i=1,2$), $ m_2/m_1=50 $, $ \delta(k_F)/(\mu_2-\mu_1)=0.02 $, $ \Gamma_2/\delta(k_F)=0.1 $, $ \Gamma_1/\delta(k_F)=0.7 $, where the electron is spinless. The unit for colorbar is $ 1/\delta(k_F) $. (b) Same as (a) but with $ \Gamma_1/\delta(k_F)=1.7 $. (c) Momentum-integrated spectral function $ A(\omega) $ for model in (a) (blue, solid) and (b) (red, dashed) and in the clean limit $ \Gamma_1=\Gamma_2=0 $ (black, dotted). The unit for $ A(\omega) $ is $ m_2 $. }
\label{fig:sp}
\end{figure}

In the general case of two distinct scattering rates $\Gamma_1 \neq \Gamma_2$, $H_0(\bk)$ and  $\Sigma$ do not commute. Then $ \gamma \neq 0$ has the nontrivial effect of altering the quasiparticle band dispersion $\epsilon_{\pm}(\bk)$, namely, damping reshapes dispersion. In particular, the quasiparticle hybridization gap at $ k=k_F $ becomes reduced, given by
\begin{numcases}{\epsilon_+(k_F) - \epsilon_-(k_F)=}
\sqrt{\delta^2(k_F)-\gamma^2}, \textrm{ when }  |\gamma|<\delta(k_F), \nonumber \\
0, \textrm{ when } |\gamma|\geq \delta(k_F).
\end{numcases}
With increasing disorder, the scattering rates $\Gamma_{1,2}$ and hence $|\gamma|$ increases. Above a critical amount of disorder $|\gamma|> \delta(k_F)$, the quasiparticle gap completely vanishes, leading to a disorder-induced semimetal. In the semimetallic phase, the quasiparticle conduction and valence bands stick together on the Fermi surface $k=k_F$, despite that the hybridization term is present. Such band sticking without fine-tuning is a remarkable and topologically robust feature which is unique to non-Hermitian band theory of finite-lifetime quasiparticles \cite{Shen2017}, but forbidden by level repulsion in Hermitian band theory. As we shall show later, quantum oscillation appears in both insulator and semimetal phases.

In Fig.~\ref{fig:sp}, we plot the spectral function $ A(\bk,\omega)$ and the density of states $ A(\omega)\equiv \int \frac{d\mathbf{k}}{(2\pi)^2}A(\mathbf{k},\omega)$ for different scattering rates $\Gamma_{1,2}$. Due to its localized nature, the $f$-orbital has a smaller  disorder-induced scattering rate $\Gamma_2 < \Gamma_1$. Panel (a) and (b) correspond to $|\gamma| <\delta(k_F)$ and $|\gamma|>\delta(k_F)$ respectively. We emphasize that the presence of two distinct scattering rates is necessary to reproduce many important features of the angle-resolved photoemission spectroscopy (ARPES) data on heavy fermion materials, which cannot be captured using $\Gamma_1= \Gamma_2$ \cite{SM}. 
We note a systematic temperature-dependent ARPES study of the Kondo insulator $\mathrm{SmB}_6$ showing that $f$-state spectra peak grows in height and narrows at low temperatures \cite{Denlinger2014}. This observation is consistent with the existence of well-defined electron quasiparticles in the zero temperature limit, but does not favor the scenario of fractional excitations.

Due to the disorder scattering, the hybridization gap is partially filled, as shown by the density of states $A(\omega)$ in Fig.~\ref{fig:sp}(c). Assuming the hybridization gap and scattering rates are small compared to the $d$-state band width, the density of states at low energy can be computed analytically \cite{SM}
\begin{equation}
A(\omega)=D_0 \mathrm{Im}\left[\frac{1}{\sqrt{\delta^2/\left[4(\omega +i\Gamma_A)^2\right]-1}}\right].
\label{eq:a}
\end{equation}
with
\begin{equation}
\Gamma_A =  \frac{m_1 \Gamma_1+ m_2\Gamma_2}{m_1+m_2}, \  \delta\equiv  \frac{2\sqrt{m_1m_2}}{m_1+m_2}\delta(k_F).
\end{equation}
Here $ m_{1,2}>0 $ are effective masses for $ d $- and $ f $-band respectively, $D_0 =D_1 + D_2$ is the total density of states from both the $d$- and $f$-bands at the Fermi energy $\omega =0$ in the absence of hybridization gap, and $\delta$ is the indirect band gap in the clean limit. The imaginary part $\Gamma_A$, a weighted sum of the two scattering rates, leads to disorder-induced broadening of density of states.
Since the $f$ band has a much larger mass $m_2 \gg m_1$, the indirect gap $\delta$ is much smaller than the hybridization  $\delta(k_F)$, and even a small scattering rate $\Gamma_2$ is sufficient to generate considerable density of states within the gap, which is consistent with previous theoretical studies \cite{Sollie1991,Riseborough2003} and experimental findings \cite{Valentine2016}. In-gap states in $\mathrm{SmB}_6$ were also reported in numerous experiments, although its origin remains an open question. For example, the low temperature electronic specific heat grows linearly with temperature $C\sim \gamma T$ instead of exponentially. Our theory is consistent with recent experiments where a variation of $\gamma$ from sample to sample is found \cite{Tan2015,Wakeham2016,PhysRevB.96.115101,Hartstein2017}. The large bulk AC conduction recently found in $\mathrm{SmB}_6$ \cite{Laurita2016} also supports the existence of localized in-gap states. 

We now show that in-gap density of states in our model exhibits quantum oscillation under magnetic field. To the leading order approximation, the scattering rates are taken to be field-independent. The density of states is then given by $ A(\omega) = -(B/\pi) \mathrm{Im} \sum_{j} \left[1/(\omega-{\cal E}_{j})\right]$,
where ${\cal E}_{j}$ denotes the complete set of complex eigenvalues of the non-Hermitian Hamiltonian with Pierels substitution ${\bf k} \rightarrow \bf k - \bf A$ ($e=\hbar=1$), i.e.,
$H(B) = H_0( {\bf k} - {\bf A}) + \Sigma$. 

For concreteness, we consider two bands with quadratic dispersion in two dimensions: $\epsilon_{i}=k^2/(2m_{i})- \mu_{i}$ ($i=1,2$), where $ m_{1,2}>0 $ are the effective masses for $ d $- and $ f $-bands respectively. We take an isotropic $p$-wave hybridization gap: $\Delta(\bk) = v (k_x s_x + k_y s_y)$. Its band structure is schematically shown in Fig.~\ref{fig:gap}.

The exact non-Hermitian Landau level spectrum of $H(B)$ is derived analytically \cite{SM}. Each Landau level $ n\geq 1 $ consists of two sets of complex eigenvalues in each spin sector denoted by $s=\uparrow, \downarrow$: 
\begin{align}
{\cal E}_{n\geq 1,\pm}^{s} &= \frac{1}{2}\Big(\epsilon_{1,n}^s-\epsilon_{2,n}^s-i\Gamma \nonumber \\
&\pm  \sqrt{\left[\left(\epsilon_{1,n}^s+\epsilon_{2,n}^s \right)-i\gamma\right]^2+v^2 (8nB)}\Big),
\label{eq:ll}
\end{align}
where $ \epsilon_{1,n}^s=B(n\pm 1/2)/m_1 +\mu_1 $ and $ \epsilon_{2,n}^s=B(n\mp 1/2)/m_2 + \mu_2$ (with upper/lower sign for $ s=\uparrow, \downarrow$). For high Landau level $ n\gg 1 $, the exact result Eq.~\eqref{eq:ll} is identical to the one obtained by simply replacing $ k\to \sqrt{2nB} $ in the zero-field dispersion (Eq.~\eqref{eq:dispersion}), and is thus also identical for both spin sectors. This agreement shows that semi-classical approximation remains valid for Landau quantization of finite-lifetime quasiparticles whose self-energy has an imaginary part.

\begin{figure}
\includegraphics[width=1\columnwidth]{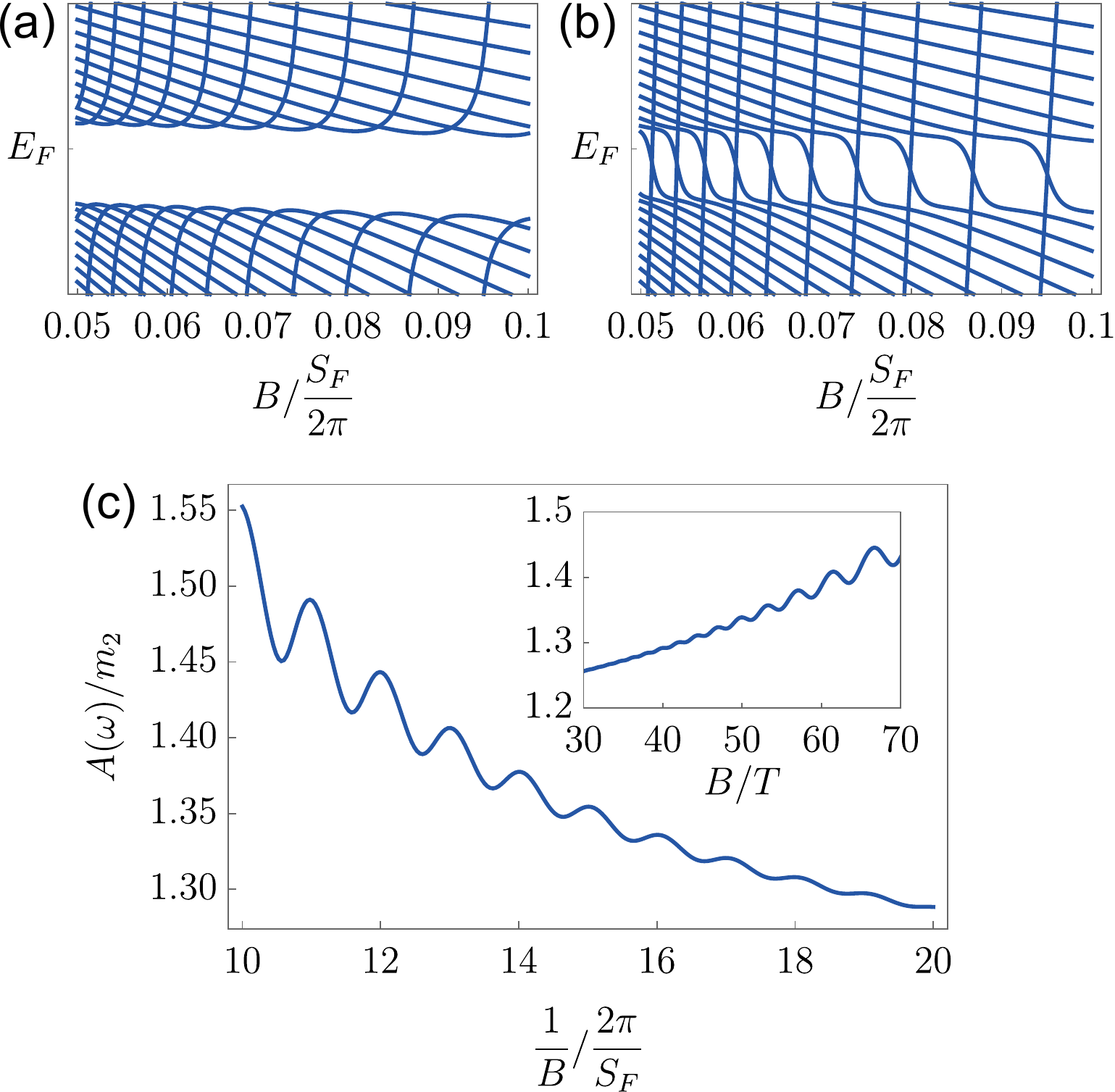}
\caption{(a) Real part of the Landau level spectrum $ \mathrm{Re}[{\cal E}_{n,\pm}^\downarrow] $ as function of magnetic field, with the same model in Fig.~\ref{fig:sp}(a). $ s=\uparrow $ sector is similar. (b) Same as (a) but with the same model in Fig.~\ref{fig:sp}(b). (c) Exact numerics of spectral function $ A(\omega=0) $ as function of $ 1/B $ and $ B $ (inset) with the model in (a). Here both spin sectors are taken into account. The unit for $ A(\omega) $ is $ m_2 $. $ m_1=m_e $ is the free electron mass, hybridization gap is $ \delta(k_F)=2\mev $. The resulting indirect band gap is $ \delta=0.56\mev $ and the oscillation frequency is $ F=800\mathrm{T} $.} 
\label{fig:qa}
\end{figure}

Typical Landau level energy spectrum is plotted as a function of the magnetic field in Fig.~\ref{fig:qa}(a) for the insulator phase, and (b) for the disorder-induced metal phase. Band edge oscillation can be seen clearly in both cases. For a given Landau level $ n $, the hybridization gap is minimized when $ B=k_F^2/(2n) $. In this way, the band edges of Landau levels oscillate with period
$ \Delta\left(1/B\right)={2}/{k_F^2}={2\pi}/{S_F} $,
where $ S_F\equiv \pi k_F^2 $ is the Fermi surface area in the absence of the hybridization. The oscillation of Landau level band edges leads to the oscillation of spectral function inside the gap, as the spectral weights inside the gap come from the tail of the broadened Landau levels. This effect, originated from the lifetime effect, persists even at zero temperature and in the limit of small (but nonzero) scattering rates.

We now turn to the field-dependent and thermally averaged density of states inside the gap, defined as $ D(\omega,T)\equiv -\int_{-\infty}^{+\infty}dE\frac{\partial n_F(E-\omega,T)}{\partial E}A(E)$, where $ n_F(\mu,T)=(e^{(E-\mu)/T}+1)^{-1} $ is the Fermi-Dirac distribution function. Under the assumption of small hybridization gap $ \delta(k_F)\ll k_F^2/\sqrt{m_1m_2}  $, weak magnetic field $ B\ll k_F^2 $ and low temperature $ T\ll \delta(k_F),B/m_{1,2} $, the density of states can be analytically computed as \cite{SM}
\begin{equation}
\begin{split}
&D(\omega=0,T)=\\
&\qquad -4\cos\left(\frac{\pi k_F^2}{B}\right)\sum_{i=\pm}\frac{M_i\frac{\pi^2 T}{\omega_{c,i}}}{\sinh\left(\frac{2\pi^2 T}{\omega_{c,i}}\right)}\exp\left(-\frac{{\cal D}_i}{B}\right),
\end{split}
\label{eq:fta}
\end{equation}
where $ \omega_{c,i}\equiv B/M_i $ are the cyclotron frequencies associated with following effective masses
\begin{equation}
M_{\pm}=\frac{(m_1 + m_2)}{2}\left[\frac{1}{\sqrt{1+\delta^2  / (4\Gamma_A^2)}}\pm \frac{m_1-m_2}{m_1 + m_2}\right],
\label{eq:lf}
\end{equation}
and $ {\cal D}_\pm $ are renormalized Dingle exponents
\begin{equation}
{\cal D}_\pm=\pi(m_1+m_2)\left( \Gamma_A \sqrt{1+\delta^2/(4\Gamma_A^2)} \pm \Gamma_D \right),
\label{eq:dingle}
\end{equation}
where $\Gamma_D \equiv (m_1\Gamma_1 -  m_2\Gamma_2)/(m_1 + m_2)$.

The analytical formula for quantum oscillation amplitude, Eq.~\eqref{eq:fta}, is one of our main results, whose form is similar to that in free electron models \cite{Champel2001} but with renormalized LK and Dingle factors. It is a sum of two oscillating components that share the same period $ \Delta(1/B)=2\pi/S_F $. This periodicity is consistent with our expectation from the Landau level spectrum shown in Fig.~\ref{fig:qa}, and has also been reported in previous works without including lifetime effects \cite{Knolle2015,Zhang2016a,Grubinskas2017}. Note that this result is not completely obvious, as the conduction band minimum and valence band maximum are located at two different momenta $k_\pm \equiv \sqrt{k_F^2\pm (m_1-m_2)\delta}$, rather than at $ k_F $ \cite{SM}. Instead of having two periods given by $ k^2_{\pm} $, the oscillation has a single period given by $ k_F^2 $, the Fermi surface area of the two bands in the absence of hybridization. We also note the phase of the oscillation is zero in Eq.~\eqref{eq:fta}, which is different from the $\pi$ phase shift in a free electron model. Under strong magnetic field $B\sim k_F^2$, there will be a field-dependent phase shift \cite{SM} as has been reported recently \cite{Qu2010,Ren2010,Wright2013,Tisserond2017}. 

The oscillation amplitude in Eq.~\eqref{eq:fta} is determined by two scattering rates $ \Gamma_A $ and $ \Gamma_D $, the indirect band gap $ \delta $ and the temperature $ T $. It reduces to familiar results in various limits. In the gapless limit $ \delta(k_F)=0 $, we reproduce the LK factors and Dingle factors for two metals. In the clean limit $\Gamma_1=\Gamma_2=0$, there is no density of state within the gap to the leading order of temperature and we find $ D(\omega=0,T)=0 $.

Although Eq.~\eqref{eq:fta} is valid both in the insulating phase $ \Gamma_{1,2}\ll \delta(k_F) $ and in the semimetallic phase $ |\gamma|>\delta(k_F) $, in the following we focus on the insulating phase, which applies to Kondo insulators. A detailed study of the disorder-induced semimetallic phase will be presented elsewhere.

We first analyze the simplest particle-hole symmetric model when $ m_1=m_2=m $ and $ \Gamma_1=\Gamma_2=\Gamma $. In this case, $ \Gamma_A=\Gamma $, $ \Gamma_D=0 $, and the LK effective mass and the Dingle exponent are
\begin{equation}
M=\frac{m}{\sqrt{1+\delta^2/(4\Gamma^2)}},\ {\cal D}=2m\pi\sqrt{\Gamma^2+\delta^2/4}.
\end{equation}
For small scattering rate $ \Gamma\ll \delta $, they reduce to $ M\sim 2m\Gamma/\delta $ and $ {\cal D}\sim m\pi\delta $. Decreasing the damping rate $ \Gamma $ leads to a smaller LK effective mass, and the Dingle exponent remains a constant controlled by the band gap.

The LK effective mass reflects the density of states inside the gap, since finite temperature effect is a thermal sampling of the spectral function through the convolution. Therefore, it is not a coincidence that Eq.~\eqref{eq:a} and Eq.~\eqref{eq:lf} look similar. Indeed, the zero field density of states inside the gap is $ A(\omega=0)= 2m\Gamma/\delta=M $, which is exactly the LK effective mass.

In the general asymmetric cases $ m_1\neq m_2 $ and $ \Gamma_1\neq \Gamma_2\ll \delta(k_F) $, the LK effective mass can vary in a wide range between $ m_1 $ and $ m_2 $ with proper choices of $ \delta $ and $ \Gamma_A $. The Dingle exponent remains a constant controlled by the band gap. This is opposite to that in normals metals when the scattering rate does not affect the LK effective mass but the Dingle factor.

As a concrete example, we present the quantum oscillation of the same model in Fig.~\ref{fig:sp}(a). The density of states as function of $ 1/B $ is shown in Fig.~\ref{fig:qa}(c). Since $ \Gamma_D\neq 0 $, the oscillation component associated with the larger Dingle factor becomes dominant \footnote{On the other hand, in the limit when $ \Gamma_D $ is very small, both LK effective masses are measurable. One may observe two plateaus in the amplitude temperature dependence as found in Ref.~\cite{Zhang2016a}. }, whose LK effective mass is $ M=8.5m_e $, in between $ m_1 $ and $ m_2 $.

We note that the band edge oscillation in the absence of the scattering rate is also reported in Ref.~\cite{Zhang2016a}. Contrary to our theory, in that case, the quantum oscillation comes from thermally excited occupation of Landau levels above the gap, hence the oscillation amplitude vanishes at zero temperature. Similar LK behavior of oscillation of thermodynamic observables in an insulator is also reported in a recent numerical study \cite{Grubinskas2017}, which is consistent with our result.

Both the LK effective mass and the Dingle factor provide testable predictions of our theory. The parameters in our analysis---the two scattering rates---can be extracted from other measurements on in-gap density of states and ARPES spectral function. The oscillation of density of states inside the gap naturally leads to magnetic susceptibility oscillation, i.e., de Haas-van Alphen effect. The in-gap states may also contribute to the quantum oscillation in resistivity, and we leave a detailed study for future work. We hope the results of this work can help understand quantum oscillations in Kondo insulators, and motivate further study of quantum oscillations from in-gap states in small-gap insulators. 

\begin{acknowledgments}
We thank Patrick Lee for insightful discussions and suggestions, Michal Papaj for collaboration on a related project, Lu Li and Yuji Matsuda for sharing their unpublished experimental results, and T. Senthil for interesting remarks. We also thank the anonymous referee who brings our attention to the enlightening experimental works. This work is supported by DOE Office of Basic Energy Sciences, Division of Materials Sciences and Engineering under Award DE-SC0010526. L.F. is partly supported by the David and Lucile Packard Foundation.
\end{acknowledgments}

\bibliography{QO_Ref}

\begin{thebibliography}{40}%
\makeatletter
\providecommand \@ifxundefined [1]{%
 \@ifx{#1\undefined}
}%
\providecommand \@ifnum [1]{%
 \ifnum #1\expandafter \@firstoftwo
 \else \expandafter \@secondoftwo
 \fi
}%
\providecommand \@ifx [1]{%
 \ifx #1\expandafter \@firstoftwo
 \else \expandafter \@secondoftwo
 \fi
}%
\providecommand \natexlab [1]{#1}%
\providecommand \enquote  [1]{``#1''}%
\providecommand \bibnamefont  [1]{#1}%
\providecommand \bibfnamefont [1]{#1}%
\providecommand \citenamefont [1]{#1}%
\providecommand \href@noop [0]{\@secondoftwo}%
\providecommand \href [0]{\begingroup \@sanitize@url \@href}%
\providecommand \@href[1]{\@@startlink{#1}\@@href}%
\providecommand \@@href[1]{\endgroup#1\@@endlink}%
\providecommand \@sanitize@url [0]{\catcode `\\12\catcode `\$12\catcode
  `\&12\catcode `\#12\catcode `\^12\catcode `\_12\catcode `\%12\relax}%
\providecommand \@@startlink[1]{}%
\providecommand \@@endlink[0]{}%
\providecommand \url  [0]{\begingroup\@sanitize@url \@url }%
\providecommand \@url [1]{\endgroup\@href {#1}{\urlprefix }}%
\providecommand \urlprefix  [0]{URL }%
\providecommand \Eprint [0]{\href }%
\providecommand \doibase [0]{http://dx.doi.org/}%
\providecommand \selectlanguage [0]{\@gobble}%
\providecommand \bibinfo  [0]{\@secondoftwo}%
\providecommand \bibfield  [0]{\@secondoftwo}%
\providecommand \translation [1]{[#1]}%
\providecommand \BibitemOpen [0]{}%
\providecommand \bibitemStop [0]{}%
\providecommand \bibitemNoStop [0]{.\EOS\space}%
\providecommand \EOS [0]{\spacefactor3000\relax}%
\providecommand \BibitemShut  [1]{\csname bibitem#1\endcsname}%
\let\auto@bib@innerbib\@empty
\bibitem [{\citenamefont {Kozii}\ and\ \citenamefont {Fu}()}]{Kozii2017}%
  \BibitemOpen
  \bibfield  {author} {\bibinfo {author} {\bibfnamefont {Vladyslav}\
  \bibnamefont {Kozii}}\ and\ \bibinfo {author} {\bibfnamefont {Liang}\
  \bibnamefont {Fu}},\ }\bibfield  {title} {\enquote {\bibinfo {title}
  {{Non-Hermitian Topological Theory of Finite-Lifetime Quasiparticles:
  Prediction of Bulk Fermi Arc Due to Exceptional Point}},}\ }\href
  {http://arxiv.org/abs/1708.05841} {\ }\Eprint
  {http://arxiv.org/abs/1708.05841} {arXiv:1708.05841} \BibitemShut {NoStop}%
\bibitem [{\citenamefont {Papaj}\ \emph {et~al.}()\citenamefont {Papaj},
  \citenamefont {Isobe},\ and\ \citenamefont {Fu}}]{Papaj2018}%
  \BibitemOpen
  \bibfield  {author} {\bibinfo {author} {\bibfnamefont {Micha{\l}}\
  \bibnamefont {Papaj}}, \bibinfo {author} {\bibfnamefont {Hiroki}\
  \bibnamefont {Isobe}}, \ and\ \bibinfo {author} {\bibfnamefont {Liang}\
  \bibnamefont {Fu}},\ }\bibfield  {title} {\enquote {\bibinfo {title} {{Bulk
  Fermi arc of disordered Dirac fermions in two dimensions}},}\ }\href
  {https://arxiv.org/abs/1802.00443} {\ }\Eprint
  {http://arxiv.org/abs/1802.00443} {arXiv:1802.00443} \BibitemShut {NoStop}%
\bibitem [{\citenamefont {Shoenberg}(1984)}]{Shoenberg1984}%
  \BibitemOpen
  \bibfield  {author} {\bibinfo {author} {\bibfnamefont {D.}~\bibnamefont
  {Shoenberg}},\ }\href {\doibase 10.1017/CBO9780511897870} {\emph {\bibinfo
  {title} {{Magnetic Oscillations in Metals}}}},\ Cambridge Monographs on
  Physics\ (\bibinfo  {publisher} {Cambridge University Press},\ \bibinfo
  {year} {1984})\BibitemShut {NoStop}%
\bibitem [{\citenamefont {Li}\ \emph {et~al.}(2014)\citenamefont {Li},
  \citenamefont {Xiang}, \citenamefont {Yu}, \citenamefont {Asaba},
  \citenamefont {Lawson}, \citenamefont {Cai}, \citenamefont {Tinsman},
  \citenamefont {Berkley}, \citenamefont {Wolgast}, \citenamefont {Eo},
  \citenamefont {Kim}, \citenamefont {Kurdak}, \citenamefont {Allen},
  \citenamefont {Sun}, \citenamefont {Chen}, \citenamefont {Wang},
  \citenamefont {Fisk},\ and\ \citenamefont {Li}}]{Li2014}%
  \BibitemOpen
  \bibfield  {author} {\bibinfo {author} {\bibfnamefont {G.}~\bibnamefont
  {Li}}, \bibinfo {author} {\bibfnamefont {Z.}~\bibnamefont {Xiang}}, \bibinfo
  {author} {\bibfnamefont {F.}~\bibnamefont {Yu}}, \bibinfo {author}
  {\bibfnamefont {T.}~\bibnamefont {Asaba}}, \bibinfo {author} {\bibfnamefont
  {B.}~\bibnamefont {Lawson}}, \bibinfo {author} {\bibfnamefont
  {P.}~\bibnamefont {Cai}}, \bibinfo {author} {\bibfnamefont {C.}~\bibnamefont
  {Tinsman}}, \bibinfo {author} {\bibfnamefont {A.}~\bibnamefont {Berkley}},
  \bibinfo {author} {\bibfnamefont {S.}~\bibnamefont {Wolgast}}, \bibinfo
  {author} {\bibfnamefont {Y.~S.}\ \bibnamefont {Eo}}, \bibinfo {author}
  {\bibfnamefont {Dae-Jeong}\ \bibnamefont {Kim}}, \bibinfo {author}
  {\bibfnamefont {C.}~\bibnamefont {Kurdak}}, \bibinfo {author} {\bibfnamefont
  {J.~W.}\ \bibnamefont {Allen}}, \bibinfo {author} {\bibfnamefont
  {K.}~\bibnamefont {Sun}}, \bibinfo {author} {\bibfnamefont {X.~H.}\
  \bibnamefont {Chen}}, \bibinfo {author} {\bibfnamefont {Y.~Y.}\ \bibnamefont
  {Wang}}, \bibinfo {author} {\bibfnamefont {Z.}~\bibnamefont {Fisk}}, \ and\
  \bibinfo {author} {\bibfnamefont {Lu}~\bibnamefont {Li}},\ }\bibfield
  {title} {\enquote {\bibinfo {title} {{Two-dimensional Fermi surfaces in Kondo
  insulator $\mathrm{SmB}_6$}},}\ }\href {\doibase 10.1126/science.1250366}
  {\bibfield  {journal} {\bibinfo  {journal} {Science}\ }\textbf {\bibinfo
  {volume} {346}},\ \bibinfo {pages} {1208--1212} (\bibinfo {year}
  {2014})}\BibitemShut {NoStop}%
\bibitem [{\citenamefont {Tan}\ \emph {et~al.}(2015)\citenamefont {Tan},
  \citenamefont {Hsu}, \citenamefont {Zeng}, \citenamefont {Hatnean},
  \citenamefont {Harrison}, \citenamefont {Zhu}, \citenamefont {Hartstein},
  \citenamefont {Kiourlappou}, \citenamefont {Srivastava}, \citenamefont
  {Johannes}, \citenamefont {Murphy}, \citenamefont {Park}, \citenamefont
  {Balicas}, \citenamefont {Lonzarich}, \citenamefont {Balakrishnan},\ and\
  \citenamefont {Sebastian}}]{Tan2015}%
  \BibitemOpen
  \bibfield  {author} {\bibinfo {author} {\bibfnamefont {B.~S.}\ \bibnamefont
  {Tan}}, \bibinfo {author} {\bibfnamefont {Y.-T.}\ \bibnamefont {Hsu}},
  \bibinfo {author} {\bibfnamefont {B.}~\bibnamefont {Zeng}}, \bibinfo {author}
  {\bibfnamefont {M.~Ciomaga}\ \bibnamefont {Hatnean}}, \bibinfo {author}
  {\bibfnamefont {N.}~\bibnamefont {Harrison}}, \bibinfo {author}
  {\bibfnamefont {Z.}~\bibnamefont {Zhu}}, \bibinfo {author} {\bibfnamefont
  {M.}~\bibnamefont {Hartstein}}, \bibinfo {author} {\bibfnamefont
  {M.}~\bibnamefont {Kiourlappou}}, \bibinfo {author} {\bibfnamefont
  {A.}~\bibnamefont {Srivastava}}, \bibinfo {author} {\bibfnamefont {M.~D.}\
  \bibnamefont {Johannes}}, \bibinfo {author} {\bibfnamefont {T.~P.}\
  \bibnamefont {Murphy}}, \bibinfo {author} {\bibfnamefont {J.-H.}\
  \bibnamefont {Park}}, \bibinfo {author} {\bibfnamefont {L.}~\bibnamefont
  {Balicas}}, \bibinfo {author} {\bibfnamefont {G.~G.}\ \bibnamefont
  {Lonzarich}}, \bibinfo {author} {\bibfnamefont {G.}~\bibnamefont
  {Balakrishnan}}, \ and\ \bibinfo {author} {\bibfnamefont {Suchitra~E.}\
  \bibnamefont {Sebastian}},\ }\bibfield  {title} {\enquote {\bibinfo {title}
  {{Unconventional Fermi surface in an insulating state}},}\ }\href {\doibase
  10.1126/science.aaa7974} {\bibfield  {journal} {\bibinfo  {journal}
  {Science}\ }\textbf {\bibinfo {volume} {349}},\ \bibinfo {pages} {287--290}
  (\bibinfo {year} {2015})}\BibitemShut {NoStop}%
\bibitem [{\citenamefont {Xiang}\ \emph {et~al.}()\citenamefont {Xiang},
  \citenamefont {Kasahara}, \citenamefont {Lawson}, \citenamefont {Asaba},
  \citenamefont {Tinsman}, \citenamefont {Chen}, \citenamefont {Sugimoto},
  \citenamefont {Kawaguchi}, \citenamefont {Sato}, \citenamefont {Li},
  \citenamefont {Yao}, \citenamefont {Chen}, \citenamefont {Iga}, \citenamefont
  {Matsuda},\ and\ \citenamefont {Li}}]{Li2017}%
  \BibitemOpen
  \bibfield  {author} {\bibinfo {author} {\bibfnamefont {Z.}~\bibnamefont
  {Xiang}}, \bibinfo {author} {\bibfnamefont {Y.}~\bibnamefont {Kasahara}},
  \bibinfo {author} {\bibfnamefont {B.}~\bibnamefont {Lawson}}, \bibinfo
  {author} {\bibfnamefont {T.}~\bibnamefont {Asaba}}, \bibinfo {author}
  {\bibfnamefont {C.}~\bibnamefont {Tinsman}}, \bibinfo {author} {\bibfnamefont
  {Lu}~\bibnamefont {Chen}}, \bibinfo {author} {\bibfnamefont {K.}~\bibnamefont
  {Sugimoto}}, \bibinfo {author} {\bibfnamefont {H.}~\bibnamefont {Kawaguchi}},
  \bibinfo {author} {\bibfnamefont {Y.}~\bibnamefont {Sato}}, \bibinfo {author}
  {\bibfnamefont {G.}~\bibnamefont {Li}}, \bibinfo {author} {\bibfnamefont
  {S.}~\bibnamefont {Yao}}, \bibinfo {author} {\bibfnamefont {Y.L.}\
  \bibnamefont {Chen}}, \bibinfo {author} {\bibfnamefont {F.}~\bibnamefont
  {Iga}}, \bibinfo {author} {\bibfnamefont {Y.}~\bibnamefont {Matsuda}}, \ and\
  \bibinfo {author} {\bibfnamefont {Lu}~\bibnamefont {Li}},\ }\bibfield
  {title} {\enquote {\bibinfo {title} {{Quantum Oscillations of Electrical
  Resistivity in an Insulator}},}\ }\href@noop {} {\bibinfo  {journal} {under
  review}\ }\BibitemShut {NoStop}%
\bibitem [{\citenamefont {Liu}\ \emph {et~al.}(2018)\citenamefont {Liu},
  \citenamefont {Hartstein}, \citenamefont {Wallace}, \citenamefont {Davies},
  \citenamefont {Hatnean}, \citenamefont {Johannes}, \citenamefont
  {Shitsevalova}, \citenamefont {Balakrishnan},\ and\ \citenamefont
  {Sebastian}}]{Hsu2018}%
  \BibitemOpen
\bibfield  {journal} {  }\bibfield  {author} {\bibinfo {author} {\bibfnamefont
  {Hsu}\ \bibnamefont {Liu}}, \bibinfo {author} {\bibfnamefont {Máté}\
  \bibnamefont {Hartstein}}, \bibinfo {author} {\bibfnamefont {Gregory~J}\
  \bibnamefont {Wallace}}, \bibinfo {author} {\bibfnamefont {Alexander~J}\
  \bibnamefont {Davies}}, \bibinfo {author} {\bibfnamefont {Monica~Ciomaga}\
  \bibnamefont {Hatnean}}, \bibinfo {author} {\bibfnamefont {Michelle~D}\
  \bibnamefont {Johannes}}, \bibinfo {author} {\bibfnamefont {Natalya}\
  \bibnamefont {Shitsevalova}}, \bibinfo {author} {\bibfnamefont {Geetha}\
  \bibnamefont {Balakrishnan}}, \ and\ \bibinfo {author} {\bibfnamefont
  {Suchitra~E}\ \bibnamefont {Sebastian}},\ }\bibfield  {title} {\enquote
  {\bibinfo {title} {Fermi surfaces in kondo insulators},}\ }\href
  {http://stacks.iop.org/0953-8984/30/i=16/a=16LT01} {\bibfield  {journal}
  {\bibinfo  {journal} {Journal of Physics: Condensed Matter}\ }\textbf
  {\bibinfo {volume} {30}},\ \bibinfo {pages} {16LT01} (\bibinfo {year}
  {2018})}\BibitemShut {NoStop}%
\bibitem [{\citenamefont {Knolle}\ and\ \citenamefont
  {Cooper}(2015)}]{Knolle2015}%
  \BibitemOpen
  \bibfield  {author} {\bibinfo {author} {\bibfnamefont {Johannes}\
  \bibnamefont {Knolle}}\ and\ \bibinfo {author} {\bibfnamefont {Nigel~R.}\
  \bibnamefont {Cooper}},\ }\bibfield  {title} {\enquote {\bibinfo {title}
  {{Quantum Oscillations without a Fermi Surface and the Anomalous de
  Haas\char21{}van Alphen Effect}},}\ }\href {\doibase
  10.1103/PhysRevLett.115.146401} {\bibfield  {journal} {\bibinfo  {journal}
  {Phys. Rev. Lett.}\ }\textbf {\bibinfo {volume} {115}},\ \bibinfo {pages}
  {146401} (\bibinfo {year} {2015})}\BibitemShut {NoStop}%
\bibitem [{\citenamefont {Knolle}\ and\ \citenamefont
  {Cooper}(2017{\natexlab{a}})}]{Knolle2017a}%
  \BibitemOpen
  \bibfield  {author} {\bibinfo {author} {\bibfnamefont {Johannes}\
  \bibnamefont {Knolle}}\ and\ \bibinfo {author} {\bibfnamefont {Nigel~R.}\
  \bibnamefont {Cooper}},\ }\bibfield  {title} {\enquote {\bibinfo {title}
  {{Anomalous de Haas\char21{}van Alphen Effect in
  $\mathrm{InAs}/\mathrm{GaSb}$ Quantum Wells}},}\ }\href {\doibase
  10.1103/PhysRevLett.118.176801} {\bibfield  {journal} {\bibinfo  {journal}
  {Phys. Rev. Lett.}\ }\textbf {\bibinfo {volume} {118}},\ \bibinfo {pages}
  {176801} (\bibinfo {year} {2017}{\natexlab{a}})}\BibitemShut {NoStop}%
\bibitem [{\citenamefont {Knolle}\ and\ \citenamefont
  {Cooper}(2017{\natexlab{b}})}]{Knolle2017}%
  \BibitemOpen
  \bibfield  {author} {\bibinfo {author} {\bibfnamefont {Johannes}\
  \bibnamefont {Knolle}}\ and\ \bibinfo {author} {\bibfnamefont {Nigel~R.}\
  \bibnamefont {Cooper}},\ }\bibfield  {title} {\enquote {\bibinfo {title}
  {{Excitons in topological Kondo insulators: Theory of thermodynamic and
  transport anomalies in $\mathrm{SmB}_{6}$}},}\ }\href {\doibase
  10.1103/PhysRevLett.118.096604} {\bibfield  {journal} {\bibinfo  {journal}
  {Phys. Rev. Lett.}\ }\textbf {\bibinfo {volume} {118}},\ \bibinfo {pages}
  {096604} (\bibinfo {year} {2017}{\natexlab{b}})}\BibitemShut {NoStop}%
\bibitem [{\citenamefont {Zhang}\ \emph {et~al.}(2016)\citenamefont {Zhang},
  \citenamefont {Song},\ and\ \citenamefont {Wang}}]{Zhang2016a}%
  \BibitemOpen
  \bibfield  {author} {\bibinfo {author} {\bibfnamefont {Long}\ \bibnamefont
  {Zhang}}, \bibinfo {author} {\bibfnamefont {Xue-Yang}\ \bibnamefont {Song}},
  \ and\ \bibinfo {author} {\bibfnamefont {Fa}~\bibnamefont {Wang}},\
  }\bibfield  {title} {\enquote {\bibinfo {title} {{Quantum Oscillation in
  Narrow-Gap Topological Insulators}},}\ }\href {\doibase
  10.1103/PhysRevLett.116.046404} {\bibfield  {journal} {\bibinfo  {journal}
  {Phys. Rev. Lett.}\ }\textbf {\bibinfo {volume} {116}},\ \bibinfo {pages}
  {046404} (\bibinfo {year} {2016})}\BibitemShut {NoStop}%
\bibitem [{\citenamefont {Baskaran}()}]{Baskaran2015}%
  \BibitemOpen
  \bibfield  {author} {\bibinfo {author} {\bibfnamefont {G.}~\bibnamefont
  {Baskaran}},\ }\bibfield  {title} {\enquote {\bibinfo {title} {{Majorana
  Fermi Sea in Insulating SmB$_{6}$: A proposal and a Theory of Quantum
  Oscillations in Kondo Insulators}},}\ }\href
  {http://arxiv.org/abs/1507.03477} {\ }\Eprint
  {http://arxiv.org/abs/1507.03477} {arXiv:1507.03477} \BibitemShut {NoStop}%
\bibitem [{\citenamefont {Erten}\ \emph {et~al.}(2017)\citenamefont {Erten},
  \citenamefont {Chang}, \citenamefont {Coleman},\ and\ \citenamefont
  {Tsvelik}}]{Erten2017}%
  \BibitemOpen
  \bibfield  {author} {\bibinfo {author} {\bibfnamefont {Onur}\ \bibnamefont
  {Erten}}, \bibinfo {author} {\bibfnamefont {Po-Yao}\ \bibnamefont {Chang}},
  \bibinfo {author} {\bibfnamefont {Piers}\ \bibnamefont {Coleman}}, \ and\
  \bibinfo {author} {\bibfnamefont {Alexei~M.}\ \bibnamefont {Tsvelik}},\
  }\bibfield  {title} {\enquote {\bibinfo {title} {{Skyrme Insulators:
  Insulators at the Brink of Superconductivity}},}\ }\href {\doibase
  10.1103/PhysRevLett.119.057603} {\bibfield  {journal} {\bibinfo  {journal}
  {Phys. Rev. Lett.}\ }\textbf {\bibinfo {volume} {119}},\ \bibinfo {pages}
  {057603} (\bibinfo {year} {2017})}\BibitemShut {NoStop}%
\bibitem [{\citenamefont {Grubinskas}\ and\ \citenamefont
  {Fritz}(2018)}]{Grubinskas2017}%
  \BibitemOpen
  \bibfield  {author} {\bibinfo {author} {\bibfnamefont {Simonas}\ \bibnamefont
  {Grubinskas}}\ and\ \bibinfo {author} {\bibfnamefont {Lars}\ \bibnamefont
  {Fritz}},\ }\bibfield  {title} {\enquote {\bibinfo {title} {{Modification of
  the Lifshitz-Kosevich formula for anomalous de Haas-van Alphen oscillations
  in inverted insulators}},}\ }\href {\doibase 10.1103/PhysRevB.97.115202}
  {\bibfield  {journal} {\bibinfo  {journal} {Phys. Rev. B}\ }\textbf {\bibinfo
  {volume} {97}},\ \bibinfo {pages} {115202} (\bibinfo {year}
  {2018})}\BibitemShut {NoStop}%
\bibitem [{\citenamefont {Sodemann}\ \emph {et~al.}(2018)\citenamefont
  {Sodemann}, \citenamefont {Chowdhury},\ and\ \citenamefont
  {Senthil}}]{Sodemann2017}%
  \BibitemOpen
  \bibfield  {author} {\bibinfo {author} {\bibfnamefont {Inti}\ \bibnamefont
  {Sodemann}}, \bibinfo {author} {\bibfnamefont {Debanjan}\ \bibnamefont
  {Chowdhury}}, \ and\ \bibinfo {author} {\bibfnamefont {T.}~\bibnamefont
  {Senthil}},\ }\bibfield  {title} {\enquote {\bibinfo {title} {{Quantum
  oscillations in insulators with neutral Fermi surfaces}},}\ }\href {\doibase
  10.1103/PhysRevB.97.045152} {\bibfield  {journal} {\bibinfo  {journal} {Phys.
  Rev. B}\ }\textbf {\bibinfo {volume} {97}},\ \bibinfo {pages} {045152}
  (\bibinfo {year} {2018})}\BibitemShut {NoStop}%
\bibitem [{\citenamefont {Pal}\ \emph {et~al.}(2016)\citenamefont {Pal},
  \citenamefont {Pi\'echon}, \citenamefont {Fuchs}, \citenamefont {Goerbig},\
  and\ \citenamefont {Montambaux}}]{Pal2017}%
  \BibitemOpen
  \bibfield  {author} {\bibinfo {author} {\bibfnamefont {Hridis~K.}\
  \bibnamefont {Pal}}, \bibinfo {author} {\bibfnamefont {Fr\'ed\'eric}\
  \bibnamefont {Pi\'echon}}, \bibinfo {author} {\bibfnamefont {Jean-No\"el}\
  \bibnamefont {Fuchs}}, \bibinfo {author} {\bibfnamefont {Mark}\ \bibnamefont
  {Goerbig}}, \ and\ \bibinfo {author} {\bibfnamefont {Gilles}\ \bibnamefont
  {Montambaux}},\ }\bibfield  {title} {\enquote {\bibinfo {title} {Chemical
  potential asymmetry and quantum oscillations in insulators},}\ }\href
  {\doibase 10.1103/PhysRevB.94.125140} {\bibfield  {journal} {\bibinfo
  {journal} {Phys. Rev. B}\ }\textbf {\bibinfo {volume} {94}},\ \bibinfo
  {pages} {125140} (\bibinfo {year} {2016})}\BibitemShut {NoStop}%
\bibitem [{\citenamefont {Pal}(2017{\natexlab{a}})}]{Pal2017a}%
  \BibitemOpen
  \bibfield  {author} {\bibinfo {author} {\bibfnamefont {Hridis~K.}\
  \bibnamefont {Pal}},\ }\bibfield  {title} {\enquote {\bibinfo {title}
  {{Quantum oscillations from inside the Fermi sea}},}\ }\href {\doibase
  10.1103/PhysRevB.95.085111} {\bibfield  {journal} {\bibinfo  {journal} {Phys.
  Rev. B}\ }\textbf {\bibinfo {volume} {95}},\ \bibinfo {pages} {085111}
  (\bibinfo {year} {2017}{\natexlab{a}})}\BibitemShut {NoStop}%
\bibitem [{\citenamefont {Pal}(2017{\natexlab{b}})}]{Pal2017b}%
  \BibitemOpen
  \bibfield  {author} {\bibinfo {author} {\bibfnamefont {Hridis~K.}\
  \bibnamefont {Pal}},\ }\bibfield  {title} {\enquote {\bibinfo {title}
  {{Unusual frequency of quantum oscillations in strongly particle-hole
  asymmetric insulators}},}\ }\href {\doibase 10.1103/PhysRevB.96.235121}
  {\bibfield  {journal} {\bibinfo  {journal} {Phys. Rev. B}\ }\textbf {\bibinfo
  {volume} {96}},\ \bibinfo {pages} {235121} (\bibinfo {year}
  {2017}{\natexlab{b}})}\BibitemShut {NoStop}%
\bibitem [{\citenamefont {Ram}\ and\ \citenamefont {Kumar}(2017)}]{Ram2017}%
  \BibitemOpen
  \bibfield  {author} {\bibinfo {author} {\bibfnamefont {Panch}\ \bibnamefont
  {Ram}}\ and\ \bibinfo {author} {\bibfnamefont {Brijesh}\ \bibnamefont
  {Kumar}},\ }\bibfield  {title} {\enquote {\bibinfo {title} {Theory of quantum
  oscillations of magnetization in kondo insulators},}\ }\href {\doibase
  10.1103/PhysRevB.96.075115} {\bibfield  {journal} {\bibinfo  {journal} {Phys.
  Rev. B}\ }\textbf {\bibinfo {volume} {96}},\ \bibinfo {pages} {075115}
  (\bibinfo {year} {2017})}\BibitemShut {NoStop}%
\bibitem [{\citenamefont {Lifshitz}\ and\ \citenamefont
  {Kosevich}(1956)}]{Lifshitz1956}%
  \BibitemOpen
  \bibfield  {author} {\bibinfo {author} {\bibfnamefont {I.~M.}\ \bibnamefont
  {Lifshitz}}\ and\ \bibinfo {author} {\bibfnamefont {A.~M.}\ \bibnamefont
  {Kosevich}},\ }\bibfield  {title} {\enquote {\bibinfo {title} {{Theory of
  Magnetic Susceptibility in Metals at Low Temperature}},}\ }\href
  {http://www.jetp.ac.ru/cgi-bin/e/index/e/2/4/p636?a=list} {\bibfield
  {journal} {\bibinfo  {journal} {Sov. Phys. JETP}\ }\textbf {\bibinfo {volume}
  {2}},\ \bibinfo {pages} {636--645} (\bibinfo {year} {1956})}\BibitemShut
  {NoStop}%
\bibitem [{\citenamefont {Champel}\ and\ \citenamefont
  {Mineev}(2001)}]{Champel2001}%
  \BibitemOpen
  \bibfield  {author} {\bibinfo {author} {\bibfnamefont {T.}~\bibnamefont
  {Champel}}\ and\ \bibinfo {author} {\bibfnamefont {V.P.}\ \bibnamefont
  {Mineev}},\ }\bibfield  {title} {\enquote {\bibinfo {title} {{de Haas–van
  Alphen effect in two- and quasi-two-dimensional metals and
  superconductors}},}\ }\href {\doibase 10.1080/13642810108216525} {\bibfield
  {journal} {\bibinfo  {journal} {Philosophical Magazine B}\ }\textbf {\bibinfo
  {volume} {81}},\ \bibinfo {pages} {55--74} (\bibinfo {year}
  {2001})}\BibitemShut {NoStop}%
\bibitem [{\citenamefont {Kishigi}\ and\ \citenamefont
  {Hasegawa}(2014)}]{Kishigi2014}%
  \BibitemOpen
  \bibfield  {author} {\bibinfo {author} {\bibfnamefont {Keita}\ \bibnamefont
  {Kishigi}}\ and\ \bibinfo {author} {\bibfnamefont {Yasumasa}\ \bibnamefont
  {Hasegawa}},\ }\bibfield  {title} {\enquote {\bibinfo {title} {{Quantum
  oscillations of magnetization in tight-binding electrons on a honeycomb
  lattice}},}\ }\href {\doibase 10.1103/PhysRevB.90.085427} {\bibfield
  {journal} {\bibinfo  {journal} {Phys. Rev. B}\ }\textbf {\bibinfo {volume}
  {90}},\ \bibinfo {pages} {085427} (\bibinfo {year} {2014})}\BibitemShut
  {NoStop}%
\bibitem [{\citenamefont {Hewson}(1993)}]{Hewson1993}%
  \BibitemOpen
  \bibfield  {author} {\bibinfo {author} {\bibfnamefont {Alexander~Cyril}\
  \bibnamefont {Hewson}},\ }\href {\doibase 10.1017/CBO9780511470752} {\emph
  {\bibinfo {title} {{The Kondo Problem to Heavy Fermions}}}},\ Cambridge
  Studies in Magnetism\ (\bibinfo  {publisher} {Cambridge University Press},\
  \bibinfo {year} {1993})\BibitemShut {NoStop}%
\bibitem [{SM()}]{SM}%
  \BibitemOpen
  \href@noop {} {}\bibinfo {note} {See Supplemental Material for calculation
  details and supplemental discussions, which includes
  Refs.\cite{Groth2009,Chen2017}.}\BibitemShut {Stop}%
\bibitem [{\citenamefont {Shen}\ \emph {et~al.}(2018)\citenamefont {Shen},
  \citenamefont {Zhen},\ and\ \citenamefont {Fu}}]{Shen2017}%
  \BibitemOpen
  \bibfield  {author} {\bibinfo {author} {\bibfnamefont {Huitao}\ \bibnamefont
  {Shen}}, \bibinfo {author} {\bibfnamefont {Bo}~\bibnamefont {Zhen}}, \ and\
  \bibinfo {author} {\bibfnamefont {Liang}\ \bibnamefont {Fu}},\ }\bibfield
  {title} {\enquote {\bibinfo {title} {{Topological Band Theory for
  Non-Hermitian Hamiltonians}},}\ }\href {\doibase
  10.1103/PhysRevLett.120.146402} {\bibfield  {journal} {\bibinfo  {journal}
  {Phys. Rev. Lett.}\ }\textbf {\bibinfo {volume} {120}},\ \bibinfo {pages}
  {146402} (\bibinfo {year} {2018})}\BibitemShut {NoStop}%
\bibitem [{\citenamefont {Denlinger}\ \emph {et~al.}(2014)\citenamefont
  {Denlinger}, \citenamefont {Allen}, \citenamefont {Kang}, \citenamefont
  {Sun}, \citenamefont {Min}, \citenamefont {Kim},\ and\ \citenamefont
  {Fisk}}]{Denlinger2014}%
  \BibitemOpen
  \bibfield  {author} {\bibinfo {author} {\bibfnamefont {Jonathan~D}\
  \bibnamefont {Denlinger}}, \bibinfo {author} {\bibfnamefont {James~W}\
  \bibnamefont {Allen}}, \bibinfo {author} {\bibfnamefont {Jeong-Soo}\
  \bibnamefont {Kang}}, \bibinfo {author} {\bibfnamefont {Kai}\ \bibnamefont
  {Sun}}, \bibinfo {author} {\bibfnamefont {Byung-II}\ \bibnamefont {Min}},
  \bibinfo {author} {\bibfnamefont {Dae-Jeong}\ \bibnamefont {Kim}}, \ and\
  \bibinfo {author} {\bibfnamefont {Zachary}\ \bibnamefont {Fisk}},\ }\bibfield
   {title} {\enquote {\bibinfo {title} {{$\mathrm{SmB}_{6}$ Photoemission: Past
  and Present}},}\ }in\ \href {\doibase doi:10.7566/JPSCP.3.017038} {\emph
  {\bibinfo {booktitle} {Proceedings of the International Conference on
  Strongly Correlated Electron Systems (SCES2013)}}},\ \bibinfo {series} {JPS
  Conference Proceedings}, Vol.~\bibinfo {volume} {3}\ (\bibinfo  {publisher}
  {Journal of the Physical Society of Japan},\ \bibinfo {year}
  {2014})\BibitemShut {NoStop}%
\bibitem [{\citenamefont {Sollie}\ and\ \citenamefont
  {Schlottmann}(1991)}]{Sollie1991}%
  \BibitemOpen
  \bibfield  {author} {\bibinfo {author} {\bibfnamefont {R.}~\bibnamefont
  {Sollie}}\ and\ \bibinfo {author} {\bibfnamefont {P.}~\bibnamefont
  {Schlottmann}},\ }\bibfield  {title} {\enquote {\bibinfo {title} {{Local
  density of states in the vicinity of a Kondo hole}},}\ }\href {\doibase
  10.1063/1.350142} {\bibfield  {journal} {\bibinfo  {journal} {Journal of
  Applied Physics}\ }\textbf {\bibinfo {volume} {70}},\ \bibinfo {pages}
  {5803--5805} (\bibinfo {year} {1991})}\BibitemShut {NoStop}%
\bibitem [{\citenamefont {Riseborough}(2003)}]{Riseborough2003}%
  \BibitemOpen
  \bibfield  {author} {\bibinfo {author} {\bibfnamefont {Peter~S.}\
  \bibnamefont {Riseborough}},\ }\bibfield  {title} {\enquote {\bibinfo {title}
  {{Collapse of the coherence gap in Kondo semiconductors}},}\ }\href {\doibase
  10.1103/PhysRevB.68.235213} {\bibfield  {journal} {\bibinfo  {journal} {Phys.
  Rev. B}\ }\textbf {\bibinfo {volume} {68}},\ \bibinfo {pages} {235213}
  (\bibinfo {year} {2003})}\BibitemShut {NoStop}%
\bibitem [{\citenamefont {Valentine}\ \emph {et~al.}(2016)\citenamefont
  {Valentine}, \citenamefont {Koohpayeh}, \citenamefont {Phelan}, \citenamefont
  {McQueen}, \citenamefont {Rosa}, \citenamefont {Fisk},\ and\ \citenamefont
  {Drichko}}]{Valentine2016}%
  \BibitemOpen
  \bibfield  {author} {\bibinfo {author} {\bibfnamefont {Michael~E.}\
  \bibnamefont {Valentine}}, \bibinfo {author} {\bibfnamefont {Seyed}\
  \bibnamefont {Koohpayeh}}, \bibinfo {author} {\bibfnamefont {W.~Adam}\
  \bibnamefont {Phelan}}, \bibinfo {author} {\bibfnamefont {Tyrel~M.}\
  \bibnamefont {McQueen}}, \bibinfo {author} {\bibfnamefont {Priscila F.~S.}\
  \bibnamefont {Rosa}}, \bibinfo {author} {\bibfnamefont {Zachary}\
  \bibnamefont {Fisk}}, \ and\ \bibinfo {author} {\bibfnamefont {Natalia}\
  \bibnamefont {Drichko}},\ }\bibfield  {title} {\enquote {\bibinfo {title}
  {{Breakdown of the Kondo insulating state in ${\mathrm{SmB}}_{6}$ by
  introducing Sm vacancies}},}\ }\href {\doibase 10.1103/PhysRevB.94.075102}
  {\bibfield  {journal} {\bibinfo  {journal} {Phys. Rev. B}\ }\textbf {\bibinfo
  {volume} {94}},\ \bibinfo {pages} {075102} (\bibinfo {year}
  {2016})}\BibitemShut {NoStop}%
\bibitem [{\citenamefont {Wakeham}\ \emph {et~al.}(2016)\citenamefont
  {Wakeham}, \citenamefont {Rosa}, \citenamefont {Wang}, \citenamefont {Kang},
  \citenamefont {Fisk}, \citenamefont {Ronning},\ and\ \citenamefont
  {Thompson}}]{Wakeham2016}%
  \BibitemOpen
  \bibfield  {author} {\bibinfo {author} {\bibfnamefont {N.}~\bibnamefont
  {Wakeham}}, \bibinfo {author} {\bibfnamefont {P.~F.~S.}\ \bibnamefont
  {Rosa}}, \bibinfo {author} {\bibfnamefont {Y.~Q.}\ \bibnamefont {Wang}},
  \bibinfo {author} {\bibfnamefont {M.}~\bibnamefont {Kang}}, \bibinfo {author}
  {\bibfnamefont {Z.}~\bibnamefont {Fisk}}, \bibinfo {author} {\bibfnamefont
  {F.}~\bibnamefont {Ronning}}, \ and\ \bibinfo {author} {\bibfnamefont
  {J.~D.}\ \bibnamefont {Thompson}},\ }\bibfield  {title} {\enquote {\bibinfo
  {title} {{Low-temperature conducting state in two candidate topological Kondo
  insulators: ${\mathrm{SmB}}_{6}$ and
  ${\mathrm{Ce}}_{3}{\mathrm{Bi}}_{4}{\mathrm{Pt}}_{3}$}},}\ }\href {\doibase
  10.1103/PhysRevB.94.035127} {\bibfield  {journal} {\bibinfo  {journal} {Phys.
  Rev. B}\ }\textbf {\bibinfo {volume} {94}},\ \bibinfo {pages} {035127}
  (\bibinfo {year} {2016})}\BibitemShut {NoStop}%
\bibitem [{\citenamefont {Orend\'a\ifmmode~\check{c}\else \v{c}\fi{}}\ \emph
  {et~al.}(2017)\citenamefont {Orend\'a\ifmmode~\check{c}\else \v{c}\fi{}},
  \citenamefont {Gab\'ani}, \citenamefont {Prist\'a\ifmmode~\check{s}\else
  \v{s}\fi{}}, \citenamefont {Ga\ifmmode~\check{z}\else \v{z}\fi{}o},
  \citenamefont {Diko}, \citenamefont {Farka\ifmmode~\check{s}\else
  \v{s}\fi{}ovsk\'y}, \citenamefont {Levchenko}, \citenamefont {Shitsevalova},\
  and\ \citenamefont {Flachbart}}]{PhysRevB.96.115101}%
  \BibitemOpen
  \bibfield  {author} {\bibinfo {author} {\bibfnamefont {Mat.}\ \bibnamefont
  {Orend\'a\ifmmode~\check{c}\else \v{c}\fi{}}}, \bibinfo {author}
  {\bibfnamefont {S.}~\bibnamefont {Gab\'ani}}, \bibinfo {author}
  {\bibfnamefont {G.}~\bibnamefont {Prist\'a\ifmmode~\check{s}\else
  \v{s}\fi{}}}, \bibinfo {author} {\bibfnamefont {E.}~\bibnamefont
  {Ga\ifmmode~\check{z}\else \v{z}\fi{}o}}, \bibinfo {author} {\bibfnamefont
  {P.}~\bibnamefont {Diko}}, \bibinfo {author} {\bibfnamefont {P.}~\bibnamefont
  {Farka\ifmmode~\check{s}\else \v{s}\fi{}ovsk\'y}}, \bibinfo {author}
  {\bibfnamefont {A.}~\bibnamefont {Levchenko}}, \bibinfo {author}
  {\bibfnamefont {N.}~\bibnamefont {Shitsevalova}}, \ and\ \bibinfo {author}
  {\bibfnamefont {K.}~\bibnamefont {Flachbart}},\ }\bibfield  {title} {\enquote
  {\bibinfo {title} {{Isosbestic points in doped $\mathrm{Sm}{\mathrm{B}}_{6}$
  as features of universality and property tuning}},}\ }\href {\doibase
  10.1103/PhysRevB.96.115101} {\bibfield  {journal} {\bibinfo  {journal} {Phys.
  Rev. B}\ }\textbf {\bibinfo {volume} {96}},\ \bibinfo {pages} {115101}
  (\bibinfo {year} {2017})}\BibitemShut {NoStop}%
\bibitem [{\citenamefont {Hartstein}\ \emph {et~al.}(2017)\citenamefont
  {Hartstein}, \citenamefont {Toews}, \citenamefont {Hsu}, \citenamefont
  {Zeng}, \citenamefont {Chen}, \citenamefont {Hatnean}, \citenamefont {Zhang},
  \citenamefont {Nakamura}, \citenamefont {Padgett}, \citenamefont
  {Rodway-Gant}, \citenamefont {Berk}, \citenamefont {Kingston}, \citenamefont
  {Zhang}, \citenamefont {Chan}, \citenamefont {Yamashita}, \citenamefont
  {Sakakibara}, \citenamefont {Takano}, \citenamefont {Park}, \citenamefont
  {Balicas}, \citenamefont {Harrison}, \citenamefont {Shitsevalova},
  \citenamefont {Balakrishnan}, \citenamefont {Lonzarich}, \citenamefont
  {Hill}, \citenamefont {Sutherland},\ and\ \citenamefont
  {Sebastian}}]{Hartstein2017}%
  \BibitemOpen
  \bibfield  {author} {\bibinfo {author} {\bibfnamefont {M}~\bibnamefont
  {Hartstein}}, \bibinfo {author} {\bibfnamefont {W~H}\ \bibnamefont {Toews}},
  \bibinfo {author} {\bibfnamefont {Y.-T.}\ \bibnamefont {Hsu}}, \bibinfo
  {author} {\bibfnamefont {B}~\bibnamefont {Zeng}}, \bibinfo {author}
  {\bibfnamefont {X}~\bibnamefont {Chen}}, \bibinfo {author} {\bibfnamefont
  {M~Ciomaga}\ \bibnamefont {Hatnean}}, \bibinfo {author} {\bibfnamefont {Q~R}\
  \bibnamefont {Zhang}}, \bibinfo {author} {\bibfnamefont {S}~\bibnamefont
  {Nakamura}}, \bibinfo {author} {\bibfnamefont {A~S}\ \bibnamefont {Padgett}},
  \bibinfo {author} {\bibfnamefont {G}~\bibnamefont {Rodway-Gant}}, \bibinfo
  {author} {\bibfnamefont {J}~\bibnamefont {Berk}}, \bibinfo {author}
  {\bibfnamefont {M~K}\ \bibnamefont {Kingston}}, \bibinfo {author}
  {\bibfnamefont {G~H}\ \bibnamefont {Zhang}}, \bibinfo {author} {\bibfnamefont
  {M~K}\ \bibnamefont {Chan}}, \bibinfo {author} {\bibfnamefont
  {S}~\bibnamefont {Yamashita}}, \bibinfo {author} {\bibfnamefont
  {T}~\bibnamefont {Sakakibara}}, \bibinfo {author} {\bibfnamefont
  {Y}~\bibnamefont {Takano}}, \bibinfo {author} {\bibfnamefont {J.-H.}\
  \bibnamefont {Park}}, \bibinfo {author} {\bibfnamefont {L}~\bibnamefont
  {Balicas}}, \bibinfo {author} {\bibfnamefont {N}~\bibnamefont {Harrison}},
  \bibinfo {author} {\bibfnamefont {N}~\bibnamefont {Shitsevalova}}, \bibinfo
  {author} {\bibfnamefont {G}~\bibnamefont {Balakrishnan}}, \bibinfo {author}
  {\bibfnamefont {G~G}\ \bibnamefont {Lonzarich}}, \bibinfo {author}
  {\bibfnamefont {R~W}\ \bibnamefont {Hill}}, \bibinfo {author} {\bibfnamefont
  {M}~\bibnamefont {Sutherland}}, \ and\ \bibinfo {author} {\bibfnamefont
  {Suchitra~E}\ \bibnamefont {Sebastian}},\ }\bibfield  {title} {\enquote
  {\bibinfo {title} {{Fermi surface in the absence of a Fermi liquid in the
  Kondo insulator $\mathrm{SmB}_6$}},}\ }\href
  {http://dx.doi.org/10.1038/nphys4295} {\bibfield  {journal} {\bibinfo
  {journal} {Nat. Phys.}\ }\textbf {\bibinfo {volume} {14}},\ \bibinfo {pages}
  {166} (\bibinfo {year} {2017})}\BibitemShut {NoStop}%
\bibitem [{\citenamefont {Laurita}\ \emph {et~al.}(2016)\citenamefont
  {Laurita}, \citenamefont {Morris}, \citenamefont {Koohpayeh}, \citenamefont
  {Rosa}, \citenamefont {Phelan}, \citenamefont {Fisk}, \citenamefont
  {McQueen},\ and\ \citenamefont {Armitage}}]{Laurita2016}%
  \BibitemOpen
  \bibfield  {author} {\bibinfo {author} {\bibfnamefont {N.~J.}\ \bibnamefont
  {Laurita}}, \bibinfo {author} {\bibfnamefont {C.~M.}\ \bibnamefont {Morris}},
  \bibinfo {author} {\bibfnamefont {S.~M.}\ \bibnamefont {Koohpayeh}}, \bibinfo
  {author} {\bibfnamefont {P.~F.~S.}\ \bibnamefont {Rosa}}, \bibinfo {author}
  {\bibfnamefont {W.~A.}\ \bibnamefont {Phelan}}, \bibinfo {author}
  {\bibfnamefont {Z.}~\bibnamefont {Fisk}}, \bibinfo {author} {\bibfnamefont
  {T.~M.}\ \bibnamefont {McQueen}}, \ and\ \bibinfo {author} {\bibfnamefont
  {N.~P.}\ \bibnamefont {Armitage}},\ }\bibfield  {title} {\enquote {\bibinfo
  {title} {{Anomalous three-dimensional bulk ac conduction within the Kondo gap
  of ${\mathrm{SmB}}_{6}$ single crystals}},}\ }\href {\doibase
  10.1103/PhysRevB.94.165154} {\bibfield  {journal} {\bibinfo  {journal} {Phys.
  Rev. B}\ }\textbf {\bibinfo {volume} {94}},\ \bibinfo {pages} {165154}
  (\bibinfo {year} {2016})}\BibitemShut {NoStop}%
\bibitem [{\citenamefont {Qu}\ \emph {et~al.}(2010)\citenamefont {Qu},
  \citenamefont {Hor}, \citenamefont {Xiong}, \citenamefont {Cava},\ and\
  \citenamefont {Ong}}]{Qu2010}%
  \BibitemOpen
  \bibfield  {author} {\bibinfo {author} {\bibfnamefont {Dong-Xia}\
  \bibnamefont {Qu}}, \bibinfo {author} {\bibfnamefont {Y.~S.}\ \bibnamefont
  {Hor}}, \bibinfo {author} {\bibfnamefont {Jun}\ \bibnamefont {Xiong}},
  \bibinfo {author} {\bibfnamefont {R.~J.}\ \bibnamefont {Cava}}, \ and\
  \bibinfo {author} {\bibfnamefont {N.~P.}\ \bibnamefont {Ong}},\ }\bibfield
  {title} {\enquote {\bibinfo {title} {{Quantum Oscillations and Hall Anomaly
  of Surface States in the Topological Insulator
  $\mathrm{Bi}_2\mathrm{Te}_3$}},}\ }\href {\doibase 10.1126/science.1189792}
  {\bibfield  {journal} {\bibinfo  {journal} {Science}\ }\textbf {\bibinfo
  {volume} {329}},\ \bibinfo {pages} {821--824} (\bibinfo {year}
  {2010})}\BibitemShut {NoStop}%
\bibitem [{\citenamefont {Ren}\ \emph {et~al.}(2010)\citenamefont {Ren},
  \citenamefont {Taskin}, \citenamefont {Sasaki}, \citenamefont {Segawa},\ and\
  \citenamefont {Ando}}]{Ren2010}%
  \BibitemOpen
  \bibfield  {author} {\bibinfo {author} {\bibfnamefont {Zhi}\ \bibnamefont
  {Ren}}, \bibinfo {author} {\bibfnamefont {A.~A.}\ \bibnamefont {Taskin}},
  \bibinfo {author} {\bibfnamefont {Satoshi}\ \bibnamefont {Sasaki}}, \bibinfo
  {author} {\bibfnamefont {Kouji}\ \bibnamefont {Segawa}}, \ and\ \bibinfo
  {author} {\bibfnamefont {Yoichi}\ \bibnamefont {Ando}},\ }\bibfield  {title}
  {\enquote {\bibinfo {title} {{Large bulk resistivity and surface quantum
  oscillations in the topological insulator
  ${\text{Bi}}_{2}{\text{Te}}_{2}\text{Se}$}},}\ }\href {\doibase
  10.1103/PhysRevB.82.241306} {\bibfield  {journal} {\bibinfo  {journal} {Phys.
  Rev. B}\ }\textbf {\bibinfo {volume} {82}},\ \bibinfo {pages} {241306}
  (\bibinfo {year} {2010})}\BibitemShut {NoStop}%
\bibitem [{\citenamefont {Wright}\ and\ \citenamefont
  {McKenzie}(2013)}]{Wright2013}%
  \BibitemOpen
  \bibfield  {author} {\bibinfo {author} {\bibfnamefont {Anthony~R.}\
  \bibnamefont {Wright}}\ and\ \bibinfo {author} {\bibfnamefont {Ross~H.}\
  \bibnamefont {McKenzie}},\ }\bibfield  {title} {\enquote {\bibinfo {title}
  {{Quantum oscillations and Berry's phase in topological insulator surface
  states with broken particle-hole symmetry}},}\ }\href {\doibase
  10.1103/PhysRevB.87.085411} {\bibfield  {journal} {\bibinfo  {journal} {Phys.
  Rev. B}\ }\textbf {\bibinfo {volume} {87}},\ \bibinfo {pages} {085411}
  (\bibinfo {year} {2013})}\BibitemShut {NoStop}%
\bibitem [{\citenamefont {Tisserond}\ \emph {et~al.}(2017)\citenamefont
  {Tisserond}, \citenamefont {Fuchs}, \citenamefont {Goerbig}, \citenamefont
  {Auban-Senzier}, \citenamefont {Mézière}, \citenamefont {Batail},
  \citenamefont {Kawasugi}, \citenamefont {Suda}, \citenamefont {Yamamoto},
  \citenamefont {Kato}, \citenamefont {Tajima},\ and\ \citenamefont
  {Monteverde}}]{Tisserond2017}%
  \BibitemOpen
  \bibfield  {author} {\bibinfo {author} {\bibfnamefont {E.}~\bibnamefont
  {Tisserond}}, \bibinfo {author} {\bibfnamefont {J.~N.}\ \bibnamefont
  {Fuchs}}, \bibinfo {author} {\bibfnamefont {M.~O.}\ \bibnamefont {Goerbig}},
  \bibinfo {author} {\bibfnamefont {P.}~\bibnamefont {Auban-Senzier}}, \bibinfo
  {author} {\bibfnamefont {C.}~\bibnamefont {Mézière}}, \bibinfo {author}
  {\bibfnamefont {P.}~\bibnamefont {Batail}}, \bibinfo {author} {\bibfnamefont
  {Y.}~\bibnamefont {Kawasugi}}, \bibinfo {author} {\bibfnamefont
  {M.}~\bibnamefont {Suda}}, \bibinfo {author} {\bibfnamefont {H.~M.}\
  \bibnamefont {Yamamoto}}, \bibinfo {author} {\bibfnamefont {R.}~\bibnamefont
  {Kato}}, \bibinfo {author} {\bibfnamefont {N.}~\bibnamefont {Tajima}}, \ and\
  \bibinfo {author} {\bibfnamefont {M.}~\bibnamefont {Monteverde}},\ }\bibfield
   {title} {\enquote {\bibinfo {title} {{Aperiodic quantum oscillations of
  particle-hole asymmetric Dirac cones}},}\ }\href
  {http://stacks.iop.org/0295-5075/119/i=6/a=67001} {\bibfield  {journal}
  {\bibinfo  {journal} {EPL (Europhysics Letters)}\ }\textbf {\bibinfo {volume}
  {119}},\ \bibinfo {pages} {67001} (\bibinfo {year} {2017})}\BibitemShut
  {NoStop}%
\bibitem [{Note1()}]{Note1}%
  \BibitemOpen
  \bibinfo {note} {On the other hand, in the limit when $ \Gamma _D $ is very
  small, both LK effective masses are measurable. One may observe two plateaus
  in the amplitude temperature dependence as found in Ref.~\cite
  {Zhang2016a}.}\BibitemShut {Stop}%
\bibitem [{\citenamefont {Groth}\ \emph {et~al.}(2009)\citenamefont {Groth},
  \citenamefont {Wimmer}, \citenamefont {Akhmerov}, \citenamefont
  {Tworzyd\l{}o},\ and\ \citenamefont {Beenakker}}]{Groth2009}%
  \BibitemOpen
  \bibfield  {author} {\bibinfo {author} {\bibfnamefont {C.~W.}\ \bibnamefont
  {Groth}}, \bibinfo {author} {\bibfnamefont {M.}~\bibnamefont {Wimmer}},
  \bibinfo {author} {\bibfnamefont {A.~R.}\ \bibnamefont {Akhmerov}}, \bibinfo
  {author} {\bibfnamefont {J.}~\bibnamefont {Tworzyd\l{}o}}, \ and\ \bibinfo
  {author} {\bibfnamefont {C.~W.~J.}\ \bibnamefont {Beenakker}},\ }\bibfield
  {title} {\enquote {\bibinfo {title} {{Theory of the Topological Anderson
  Insulator}},}\ }\href {\doibase 10.1103/PhysRevLett.103.196805} {\bibfield
  {journal} {\bibinfo  {journal} {Phys. Rev. Lett.}\ }\textbf {\bibinfo
  {volume} {103}},\ \bibinfo {pages} {196805} (\bibinfo {year}
  {2009})}\BibitemShut {NoStop}%
\bibitem [{\citenamefont {Chen}\ \emph {et~al.}(2017)\citenamefont {Chen},
  \citenamefont {Xu}, \citenamefont {Niu}, \citenamefont {Jiang}, \citenamefont
  {Peng}, \citenamefont {Xu}, \citenamefont {Wen}, \citenamefont {Ding},
  \citenamefont {Huang}, \citenamefont {Shu}, \citenamefont {Zhang},
  \citenamefont {Lee}, \citenamefont {Strocov}, \citenamefont {Shi},
  \citenamefont {Bisti}, \citenamefont {Schmitt}, \citenamefont {Huang},
  \citenamefont {Dudin}, \citenamefont {Lai}, \citenamefont {Kirchner},
  \citenamefont {Yuan},\ and\ \citenamefont {Feng}}]{Chen2017}%
  \BibitemOpen
  \bibfield  {author} {\bibinfo {author} {\bibfnamefont {Q.~Y.}\ \bibnamefont
  {Chen}}, \bibinfo {author} {\bibfnamefont {D.~F.}\ \bibnamefont {Xu}},
  \bibinfo {author} {\bibfnamefont {X.~H.}\ \bibnamefont {Niu}}, \bibinfo
  {author} {\bibfnamefont {J.}~\bibnamefont {Jiang}}, \bibinfo {author}
  {\bibfnamefont {R.}~\bibnamefont {Peng}}, \bibinfo {author} {\bibfnamefont
  {H.~C.}\ \bibnamefont {Xu}}, \bibinfo {author} {\bibfnamefont {C.~H.~P.}\
  \bibnamefont {Wen}}, \bibinfo {author} {\bibfnamefont {Z.~F.}\ \bibnamefont
  {Ding}}, \bibinfo {author} {\bibfnamefont {K.}~\bibnamefont {Huang}},
  \bibinfo {author} {\bibfnamefont {L.}~\bibnamefont {Shu}}, \bibinfo {author}
  {\bibfnamefont {Y.~J.}\ \bibnamefont {Zhang}}, \bibinfo {author}
  {\bibfnamefont {H.}~\bibnamefont {Lee}}, \bibinfo {author} {\bibfnamefont
  {V.~N.}\ \bibnamefont {Strocov}}, \bibinfo {author} {\bibfnamefont
  {M.}~\bibnamefont {Shi}}, \bibinfo {author} {\bibfnamefont {F.}~\bibnamefont
  {Bisti}}, \bibinfo {author} {\bibfnamefont {T.}~\bibnamefont {Schmitt}},
  \bibinfo {author} {\bibfnamefont {Y.~B.}\ \bibnamefont {Huang}}, \bibinfo
  {author} {\bibfnamefont {P.}~\bibnamefont {Dudin}}, \bibinfo {author}
  {\bibfnamefont {X.~C.}\ \bibnamefont {Lai}}, \bibinfo {author} {\bibfnamefont
  {S.}~\bibnamefont {Kirchner}}, \bibinfo {author} {\bibfnamefont {H.~Q.}\
  \bibnamefont {Yuan}}, \ and\ \bibinfo {author} {\bibfnamefont {D.~L.}\
  \bibnamefont {Feng}},\ }\bibfield  {title} {\enquote {\bibinfo {title}
  {{Direct observation of how the heavy-fermion state develops in
  ${\mathrm{CeCoIn}}_{5}$}},}\ }\href {\doibase 10.1103/PhysRevB.96.045107}
  {\bibfield  {journal} {\bibinfo  {journal} {Phys. Rev. B}\ }\textbf {\bibinfo
  {volume} {96}},\ \bibinfo {pages} {045107} (\bibinfo {year}
  {2017})}\BibitemShut {NoStop}%
\end{thebibliography}%

\widetext
\clearpage


\section*{Supplementary material}

# Supplemental Material for "Quantum Oscillation from In-gap States and Non-Hermitian Landau Level Problem"

Huitao Shen[1] and Liang Fu[1]

[1]Department of Physics, Massachusetts Institute of Technology, Cambridge, Massachusetts 02139, USA

## CONTENTS

In this Supplemental Material, we present calculation details and supplement discussions. Although we will focus on a special case of the generic two-band model (Eq. (1) in the main text), many conclusions are governed only by the low energy behavior of the model and are thus general.

# I. MODEL

## A. Model Hamiltonian

Consider the two-band model

$$H_0(\mathbf{k}) = \begin{pmatrix} \epsilon_1(k) & \Delta(\mathbf{k}) \\ \Delta(\mathbf{k}) & -\epsilon_2(k) \end{pmatrix},$$  (1)

where the diagonal term is given by two quadratic bands in two dimensions

$$\epsilon_1 = \frac{k^2}{2m_1} + \frac{g}{2} + \mu_0, \ \epsilon_2 = \frac{k^2}{2m_2} + \frac{g}{2} - \mu_0.$$  (2)

Here $k \equiv \sqrt{k_x^2 + k_y^2}$, $m_i > 0$ for $i = 1, 2$ are the effective masses for the two bands, $g$ is the fundamental gap of the two bands at $k = 0$, and $\mu_0$ is the constant energy shift for later convenience. $g < 0$ represents the inverted band regime, which is of our interest in this paper. For convenience, we define $1/m_\pm \equiv 1/m_1 \pm 1/m_2$. Without hybridization $\Delta(\mathbf{k}) = 0$, the two bands touch at $k_F \equiv \sqrt{-2m_+ g}$. $\mu_0 \equiv -k_F^2/(4m_-)$ so that $\epsilon_1(k_F) = \epsilon_2(k_F) = 0$.

As in the main text, we take an isotropic $p$-wave hybridization gap: $\Delta(\mathbf{k}) = v(k_x s_x + k_y s_y)$. This Hamiltonian commutes with $\sigma_z s_z$ and hence is a sum of two Blocks: $H_0 = H^\uparrow \oplus H^\downarrow$, where $H^{\uparrow,\downarrow}$ is $2 \times 2$ matrix associated with spin sector $\sigma_z s_z = \pm 1$, whose off-diagonal element is $v(k_x \pm ik_y)$ respectively. Since $H^\uparrow = (H^\downarrow)^T$, all the results for $H^\downarrow$ can be easily transferred to $H^\uparrow$. In the following, we will only consider $H^\downarrow$, which can be seen as a $2 \times 2$ Hamiltonian with $\Delta(\mathbf{k}) = v(k_x - ik_y)$.

## B. Hybridzation Gap and Band Gap

Before proceeding further, we would like to remark on the "gaps" involved in this model. The band dispersion of the Bloch Hamiltonian given by Eq. (1) and (2) is

$$E_\pm(k) = \frac{1}{2} \left[ \frac{k^2 - k_F^2}{2m_-} \pm \sqrt{\left( \frac{k^2 - k_F^2}{2m_+} \right)^2 + \left( \frac{k\delta(k_F)}{k_F} \right)^2} \right],$$  (3)

where $\delta(k_F) \equiv 2\sqrt{-2m_+ v^2 g}$ is the energy difference of the two bands at $k = k_F$. $\delta(k_F)$ is also understood as the "hybridization gap". However, in the general case when $m_1 \neq m_2$, the band gap is in fact indirect. By finding the solution of $\partial E_\pm/\partial k = 0$, the conduction band bottom and the valence band top are located at

$$k_\pm^2 = k_F^2 \pm \delta(k_F) \sqrt{m_1 m_2} \frac{m_1 - m_2}{m_1 + m_2}.$$  (4)

Here we assume $\delta(k_F) \ll k_F^2/m_+$ and take $m_1 < m_2$ without loss of generality. The indirect band gap is then

$$\delta \equiv E_+(k_+) - E_-(k_-) = \frac{2\sqrt{m_1 m_2}}{m_1 + m_2} \delta(k_F).$$  (5)

In this way, Equation (4) can also be expressed with the indirect band gap: $k_\pm^2 = k_F^2 \pm (m_1 - m_2)\delta/2$.

When $m_2/m_1 = 50$, we have $\delta/\delta(k_F) = 0.28$. This is nearly 4 times difference. A schematic of the hybridization gap and the indirect band gap is shown in Fig. 1 in the main text.

# II. ESTIMATING ELECTRON'S SELF-ENERGY

Here we follow the disorder Hamiltonian $H = \sum_{\mathbf{k}} c_{\mathbf{k}}^\dagger H_0(\mathbf{k}) c_{\mathbf{k}} + \int d\mathbf{r} U(\mathbf{r}) c_{\mathbf{r}}^\dagger \Lambda c_{\mathbf{r}}$, presented in the main text. To recapitulate, $c^\dagger \equiv (d^\dagger, f^\dagger)$ is the electron creation operator for the two orbitals. $U(\mathbf{r})$ is the impurity potential, which

is assumed to be short-ranged and characterized by $\langle U(\mathbf{r}) \rangle = 0$ and $\langle U(\mathbf{r}) U(\mathbf{r}') \rangle = n_{\text{imp}} U_0^2 \delta(\mathbf{r} - \mathbf{r}')$ under disorder average $\langle \dots \rangle$, where $n_{\text{imp}}$ is the impurity density. $\Lambda$ is the electron-impurity scattering vertex of the form $\Lambda = \alpha I + \beta \sigma_z$.

To estimate electron's self-energy, we will consider self-consistent Born approximation (SCBA) and self-consistent T-matrix approximation (SCTA) in the following. T-matrix for a single isolated impurity is computed as $T = U + UGT$, and the impurity averaged electron's self-energy is given by $\Sigma = n_{\text{imp}} T$. In SCBA, only the expansion to the second order of $U$ is considered and by imposing self-consistency condition we essentially evaluate all "rainbow" diagrams (Fig. 1(a)). In SCTA, expansion of all orders are considered and by imposing self-consistency condition we essentially evaluate all "non-crossing" diagrams (Fig. 1(b)).

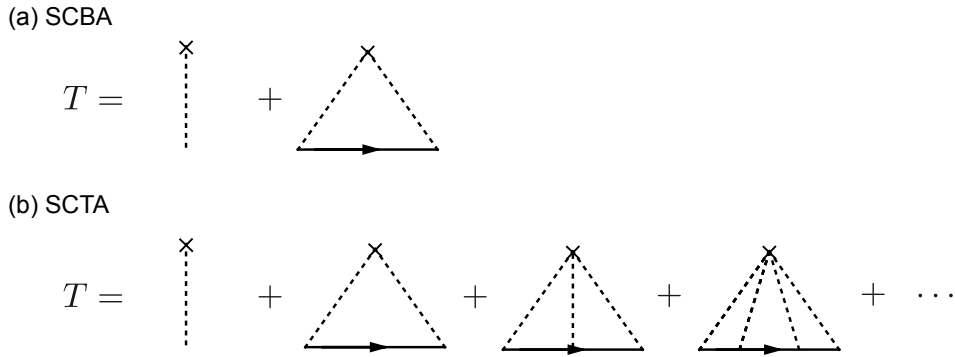

FIG. 1. Feynman diagrams for single impurity T-matrix in SCBA and SCTA. Here X, dashed line and solid line represents impurity, impurity potential and *full* fermion Green's function respectively.

### A. Self-consistent Born Approximation

#### 1. Conditions for Diagonal Self-energy

Since we are interested in the diagonal self-energy matrix $\Sigma$, we first briefly comment on the conditions for diagonal electron's self-energy $\Sigma$. There are two scenarios when solutions with diagonal $\Sigma$ exist.

First scenario is the $p$-wave hybridization where $\Delta(\mathbf{k}) = -\Delta(-\mathbf{k})$. In this case, the self-consistent equation in SCBA is

$$\Sigma(\omega) = n_{\text{imp}} U_0 + n_{\text{imp}} U_0^2 \int \frac{d^2\mathbf{k}}{(2\pi)^2} \Lambda (\omega - H_0 - \Sigma(\omega))^{-1} \Lambda \tag{6}$$

$$= n_{\text{imp}} U_0 + n_{\text{imp}} U_0^2 \int \frac{d^2\mathbf{k}}{(2\pi)^2} \frac{1}{z_1 z_2 - \Delta^2(\mathbf{k})} \begin{pmatrix} (\alpha+\beta)^2 z_2 & (\alpha^2-\beta^2)\Delta(\mathbf{k}) \\ (\alpha^2-\beta^2)\Delta(\mathbf{k}) & (\alpha+\beta)^2 z_1 \end{pmatrix}, \tag{7}$$

where

$$z_1 = \omega - \Sigma_1 - \frac{k^2}{2m_1} - \frac{g}{2}, \ z_2 = \omega - \Sigma_2 + \frac{k^2}{2m_2} + \frac{g}{2}, \tag{8}$$

are both functions even in $\mathbf{k}$. The odd function nature of $\Delta(\mathbf{k})$ guarantees the off-diagonal term in $\Sigma$ to vanish after the integration.

Another scenario is the "incoherent" $s$-wave hybridization $\Delta(\mathbf{k}) = \Delta(-\mathbf{k})$, where there are two different impurity potentials $U_1$ and $U_2$ that scatter incoherently with two orbitals. It is based on the observation that if the impurity potential is only on one orbital, i.e., $\Lambda = (I \pm \sigma_z)/2$ or $\alpha = \beta$, the off-diagonal term in $\Sigma$ will not be generated even for $s$-wave hybridization (Eq. (7)). To be more concrete, let

$$U(\mathbf{r})\Lambda = \begin{pmatrix} U_1(\mathbf{r}) & 0 \\ 0 & U_2(\mathbf{r}) \end{pmatrix}, \tag{9}$$

with $\langle U_i(\mathbf{r}) \rangle_j = 0$, $\langle U_i(\mathbf{r}) U_i(\mathbf{r}') \rangle_j = n_i U_i^2 \delta(\mathbf{r} - \mathbf{r}') \delta_{ij}$ for $i, j = 1, 2$. Here $\langle \cdots \rangle_i$ is the disorder average under impurity potential $i$. In this case, the SCBA equation becomes

$$\Sigma(\omega) = n_1 U_1^2 \int \frac{d^2\mathbf{k}}{(2\pi)^2} \Lambda_1 (\omega - H_0 - \Sigma(\omega))^{-1} \Lambda_1 + n_2 U_2^2 \int \frac{d^2\mathbf{k}}{(2\pi)^2} \Lambda_2 (\omega - H_0 - \Sigma(\omega))^{-1} \Lambda_2, \tag{10}$$

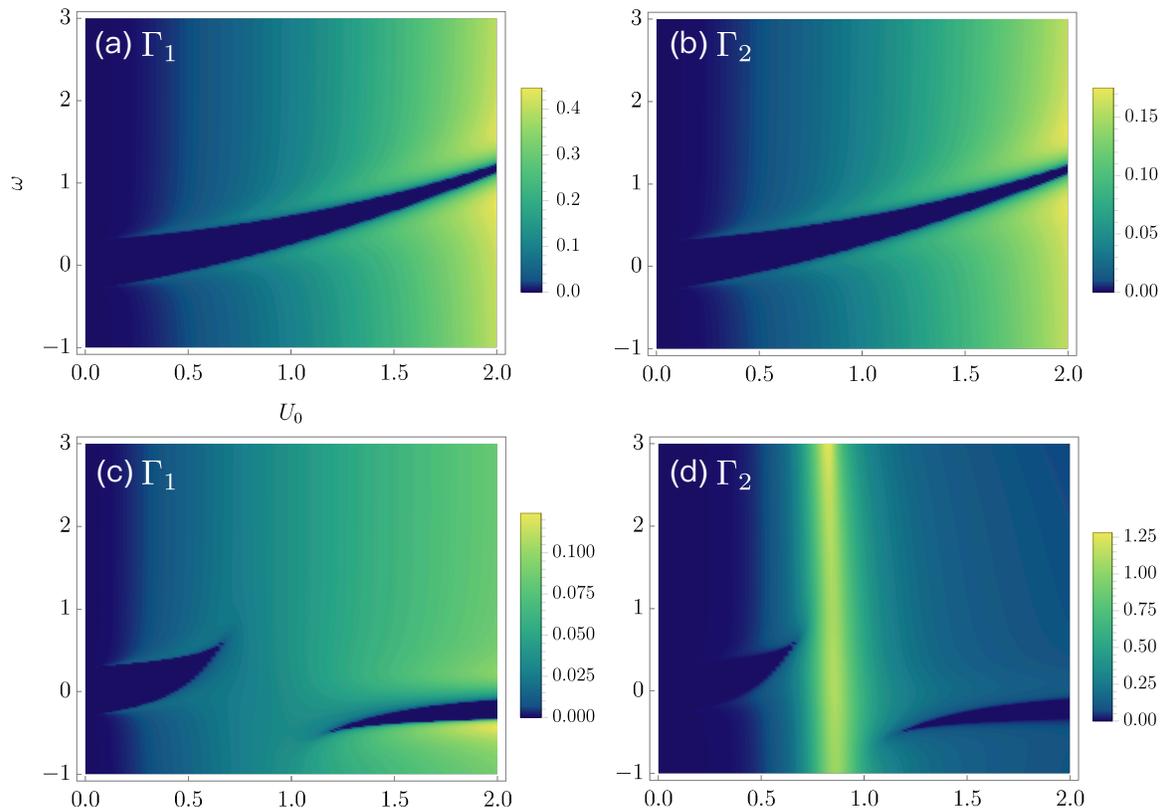

FIG. 2. Numerical solution of (a)(b) SCBA (Eq. (7)), (c)(d) SCTA (Eq. (13)), with $n_{\text{imp}}/(m_1\delta(k_F)) = 5$, $m_2/m_1 = 10$ and $-\delta(k_F)/g = 0.02$. The impurity scattering vertex is $\Gamma = \alpha I + \beta\sigma_z$ with $\alpha = 0.6$ and $\beta = 0.4$. The cutoff is chosen so that $k_{\max} = 10^4 k_F$. The unit for $U_0$ is $1/m_1$. All energy units are $\delta(k_F)$.

where $\Lambda_{1,2} = (I \pm \sigma_z)/2$. It is easy to verify both integrals in Eq. (10) is proportional to $\Lambda_{1,2} = (I \pm \sigma_z)/2$ respectively, so that the resulting $\Sigma$ is diagonal.

### 2. Numerical Solutions

The numerical solution to Eq. (7) for $p$-wave hybridization $\Delta(\mathbf{k}) = v(k_x - ik_y)$ is shown in Fig. 2(a)(b). The solution to "incoherent" $s$-wave hybridization is quantitatively the same and will thus be omitted. For this continuum model in two dimensions, the integral in Eq. (7) diverges logarithmically. In order to obtain reasonable results, we take a large cutoff in momentum $k_{\max}$ that is much larger than all momentum scales in the problem. We confirm all following results are qualitatively independent of the cutoff.

The self-energy $\Sigma$ is decomposed as

$$\text{Re}\Sigma = \begin{pmatrix} \bar{\mu} + \frac{\bar{g}}{2} & 0 \\ 0 & \bar{\mu} - \frac{\bar{g}}{2} \end{pmatrix}, \ \text{Im}\Sigma = \begin{pmatrix} \Gamma_1 & 0 \\ 0 & \Gamma_2 \end{pmatrix}, \tag{11}$$

where $\bar{\mu}$ and $\bar{g}$ are the renormalization of the chemical potential and fundamental gap, and $\Gamma_i$ is the scattering rate for band $\epsilon_i$.

The normalization of the chemical potential and the fundamental gap is always nonzero for nonzero $U_0$, and is weakly dependent on $\omega$. We find that $\bar{g}$ always tends to invert the band gap further (not shown in the figure), which is consistent with the theory of topological Anderson insulators [1].

The scattering rates of the two bands $\Gamma_1$ and $\Gamma_2$ are strictly zero for $\omega$ inside the gap and small $U_0$. This is because for small impurity potential $n_{\text{imp}}U_0 \ll \delta(k_F)$, the conservation of energy cannot be satisfied in order for electrons in the noninteracting bands to be scattered into the gap. In other words, the mechanism that generates the in-gap quasiparticle is non-perturbative in the hybridization gap $\delta(k_F)$. For large enough scattering potential $n_{\text{imp}}U_0 \sim \delta(k_F)$, $\Gamma_1(\omega)$ and $\Gamma_2(\omega)$ can be treated as nonzero constants at low energy $|\omega| \lesssim \delta(k_F)$. This is the assumption for the calculations in the rest of this paper.

## B. Self-consistent T-Matrix Approximation

The self-consistent equation in SCTA is

$$\Sigma(\omega) = n_{\mathrm{imp}}U_0 + n_{\mathrm{imp}}U_0 \int \frac{d^2\mathbf{k}}{(2\pi)^2} \Lambda(\omega - H_0 - \Sigma(\omega))^{-1}\Sigma \tag{12}$$

$$= n_{\mathrm{imp}}U_0 + n_{\mathrm{imp}}U_0^2 \int \frac{d^2\mathbf{k}}{(2\pi)^2} \frac{1}{z_1 z_2 - \Delta^2(\mathbf{k})} \begin{pmatrix} (\alpha+\beta)\Sigma_1 z_2 & (\alpha+\beta)\Sigma_2\Delta(\mathbf{k}) \\ (\alpha-\beta)\Sigma_1\Delta(\mathbf{k}) & (\alpha+\beta)\Sigma_2 z_1 \end{pmatrix}. \tag{13}$$

It is not hard to see the condition for diagonal self-energy for SCBA and SCTA are the same.

The numerical solution to Eq. (13) for $p$-wave hybridization $\Delta(\mathbf{k}) = v(k_x - ik_y)$ is shown in Fig. 2(c)(d). For weak disorder potential, the results for SCBA and SCTA are similar. For strong disorder potential, there are more features in SCTA, such as the non-monotonous behavior of $\Gamma_2$ with increasing disorder potential $U_0$. Nevertheless, when the scattering potential $n_{\mathrm{imp}}U_0 \sim \delta(k_F)$, $\Gamma_1(\omega)$ and $\Gamma_2(\omega)$ can be still treated as nonzero constants at low energy $|\omega| \lesssim \delta(k_F)$ in SCTA.

## III. MOMENTUM RESOLVED SPECTRAL FUNCTION AT ZERO FIELD

Here we analyze momentum resolved spectral functions $A(\mathbf{k}, \omega)$ in detail by comparing with the ARPES data from the experiment. Despite the simplicity of our model, the spectral function in Fig. 3(a) compares well with recent high-resolution angle-resolved photoemission spectroscopy (ARPES) data on the heavy fermion material CeInCo$_5$ taken at low temperatures [2]. Although the band structure of Kondo insulators may in general be different from that of CeInCo$_5$. We use the example of this material to show the relevance of our simple model (Eq. (2) and (1)) to heavy fermion materials, where typically a heavy $f$-band, a light $d$-band and the hybridization between them are involved.

We emphasize that the presence of two distinct scattering rates is necessary to reproduce many important features of the ARPES data, which cannot be captured using $\Gamma_1 = \Gamma_2$, which we will show in the following.

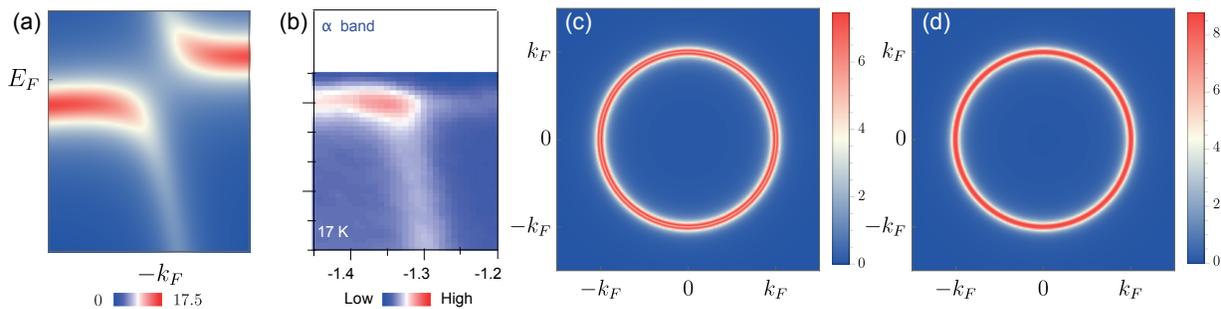

FIG. 3. (a) Same as Fig. 2(a) in the main text. (b)ARPES data of CeCoIn$_5$ taken from FIG. 6 in Ref. [2]. (c)(d)$A(\mathbf{k}, E_F)$ as function of $\mathbf{k}$. Here $m_2/m_1 = 50$, $-\delta(k_F)/g = 0.01$ and $\Gamma_2/\delta(k_F) = 0.1$. (c)$\Gamma_1/\delta(k_F) = 0.7$; (d) $\Gamma_1/\delta(k_F) = 1.7$. The unit for $A(\mathbf{k}, E_F)$ is $1/\delta(k_F)$.

We first plot $A(\mathbf{k}, E_F)$ in momentum space. One can clearly see the Fermi surface of quasiparticles inside the hybridization gap. The two Fermi surfaces appeared in Fig. 3(a) are due to the indirect band gap of our model.

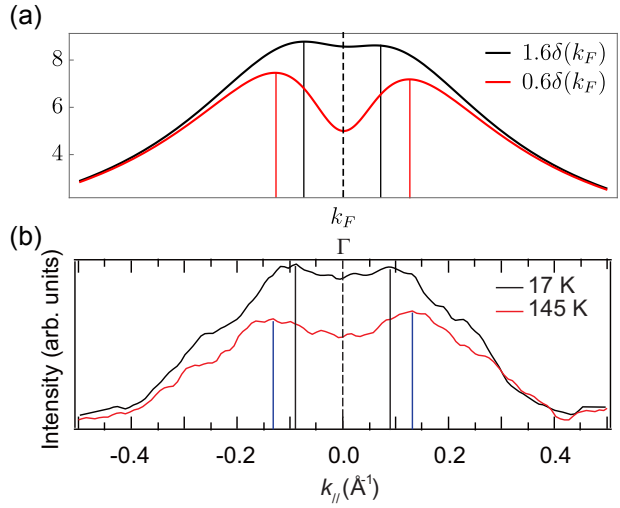

FIG. 4. (a)$A(k, E_F)$ as function of $k$ for the same models in Fig. 3, with black line being $\gamma = 1.6\delta(k_F)$ ($\Gamma_2/\delta(k_F) = 1.6$, red line being $\gamma = 0.6\delta(k_F)$ ($\Gamma_2/\delta(k_F) = 0.7$). The unit for $A(k, E_F)$ is $1/\delta(k_F)$. (b) MDC ARPES result taken from FIG. 7(d) in Ref. [2].

To examine the momentum dependence more carefully, we plot $A(k, E_F)$ near $k = k_F$ in Fig. 4(a). This corresponds to the momentum distribution curve (MDC) in the ARPES data. With increasing $|\gamma|$, two features become evident: i) the spectral densities are more concentrated near the hybridzatin gap; ii) the two peaks of the spectral function become closer. We compare the actual MDC from the experiment in Fig. 4(b). There are indeed two peaks in the the MDC data, which are consistent with the two Fermi surfaces in our model. More interestingly, the temperature dependence of the MDC data can be fit well by tuning $\gamma$. In this way, $|\gamma|$ increases with lowering the temperature.

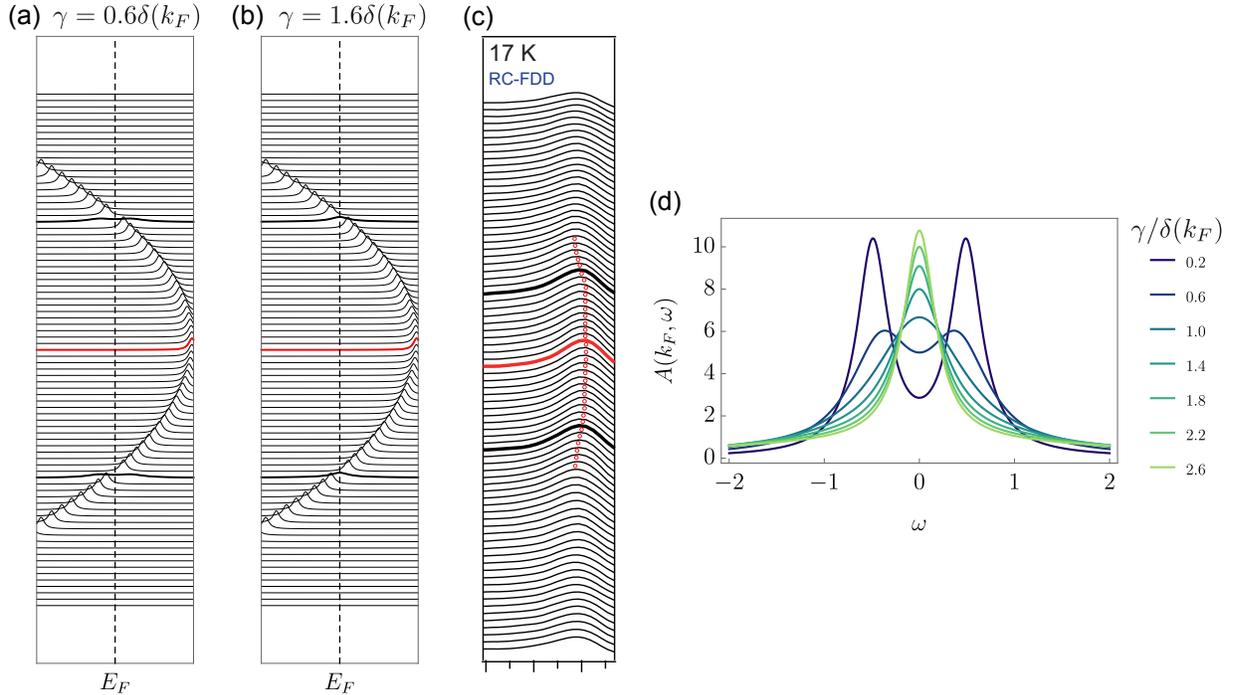

FIG. 5. (a)(b)$A(k, \omega)$ as function of both $k$ and $\omega$ for the same models in Fig. 3. The red line represents $k = 0$ and the black lines represent $k = \pm k_F$. (c) EDC ARPES result taken from FIG. 7(b) in Ref. [2]. (d) $A(k_F, \omega)$ as function of $\omega$ for the same model in Fig. 3 but with different $\Gamma_1$ hence $\gamma$. The unit for spectral function is $1/\delta(k_F)$.

Finally, we plot $A(\mathbf{k}, \omega)$ as function of $\omega$ in both Fig. 5. This corresponds to the energy distribution curve (EDC) in the ARPES data. Here we focus on the feature at $k = k_F$. The positions of the peaks represent the real part of

the poles of the Green's function. When $|\gamma| \geq \delta(k_F)$, the two peaks merge into one at $\omega = E_F$, representing the vanishing of the hybridization gap and the disorder-induced metal. We compare the actual MDC from the experiment in Fig. 5(c). The single peak at $k = k_F$ suggests that $|\gamma| \gtrsim \delta(k_F)$, i.e., the system is in or at least very near the disorder-induced metal phase.

## IV. LOW ENERGY DENSITY OF STATES AT ZERO FIELD

Consider the following quasiparticle Hamiltonian $H(k) = H_0(k) + \Sigma$ where $H_0(\mathbf{k})$ is given by Eq. (1) and Eq. (2) and $\Sigma$ has constant diagonal imaginary part

$$\Sigma = \begin{pmatrix} -i\Gamma_1 & 0 \\ 0 & -i\Gamma_2 \end{pmatrix}. \tag{14}$$

By directly diagonalizing the matrix, we obtain the spectrum for the quasiparticle Hamiltonian:

$$\mathcal{E}_{\pm}(k) = \frac{1}{2}\left[ \frac{k^2 - k_F^2}{2m_-} - i\Gamma \pm \sqrt{\left( \frac{k^2 - k_F^2}{2m_+} - i\gamma \right)^2 + \left( \frac{k\delta(k_F)}{k_F} \right)^2} \right], \tag{15}$$

where $\Gamma \equiv \Gamma_1 + \Gamma_2$, $\gamma \equiv \Gamma_1 - \Gamma_2$. Note the similarity between Eq. (3) and (15).

The corresponding spectral function is

$$A(\mathbf{k}, \omega) = -2\text{Im}\left( \frac{1}{\omega - \mathcal{E}_+(\mathbf{k})} + \frac{1}{\omega - \mathcal{E}_-(\mathbf{k})} \right) = -4\text{Im}\frac{\omega - \frac{1}{2}\left( \frac{k^2 - k_F^2}{2m_-} - i\Gamma \right)}{\left[ \omega - \frac{1}{2}\left( \frac{k^2 - k_F^2}{2m_-} - i\Gamma \right) \right]^2 - \left( \frac{k^2 - k_F^2}{2m_+} - i\gamma \right)^2 - \left( \frac{k\delta(k_F)}{k_F} \right)^2}. \tag{16}$$

The density of states can be computed as the momentum integral of the spectral function:

$$A(\omega) = \int d\mathbf{k}A(\mathbf{k}, \omega) = \pi \int dk^2 A(k, \omega) \tag{17}$$

$$= -4\pi\text{Im}\int_0^{+\infty} dq \frac{\omega - \frac{1}{2}\left( \frac{q - k_F^2}{2m_-} - i\Gamma \right)}{\left[ \omega - \frac{1}{2}\left( \frac{q - k_F^2}{2m_-} - i\Gamma \right) \right]^2 - \left( \frac{q - k_F^2}{2m_+} - i\gamma \right)^2 - q\left( \frac{\delta(k_F)}{k_F} \right)^2} \tag{18}$$

$$= -4\pi\text{Im}\int_0^{+\infty} dq \frac{aq + b}{q^2 + cq + d}, \tag{19}$$

$$= -4\pi\text{Im}\left[ \frac{2b - ac}{\sqrt{4d - c^2}} \arctan\left( \frac{2q + c}{\sqrt{4d - c^2}} \right) + \frac{a}{2}\log(q^2 + cq + d) \right] \Big|_0^{+\infty}, \tag{20}$$

with

$$a \equiv -m_1 + m_2, \tag{21}$$

$$b \equiv k_F^2(m_1 - m_2) - 2m_1 m_2(2\omega + i\Gamma), \tag{22}$$

$$c \equiv -2k_F^2 - 2m_1(\omega + i\Gamma_1) + 2m_2(\omega + i\Gamma_2) + m_1 m_2 \delta^2(k_F)/k_F^2, \tag{23}$$

$$d \equiv \left[ k_F^2 + 2m_1(\omega + i\Gamma_1) \right] \left[ k_F^2 - 2m_2(\omega + i\Gamma_2) \right]. \tag{24}$$

From Eq. (17) to (18) we have used the substitution $q = k^2$. Under the assumption of small hybridization gap $\delta(k_F) \ll k_F^2/\sqrt{m_1 m_2}$, we can expand relevant expressions to the leading order of $k_F$:

$$2b - ac = -2(m_1 + m_2)\left[ m_1(\omega + i\Gamma_1) + m_2(\omega + i\Gamma_2) \right], \tag{25}$$

$$4d - c^2 = 4m_1 m_2 \delta^2(k_F) - 4\left[ m_1(\omega + i\Gamma_1) + m_2(\omega + i\Gamma_2) \right]^2. \tag{26}$$

On the other hand,

$$\arctan\left( \frac{2q + c}{\sqrt{4d - c^2}} \right) \Big|_{q \to +\infty} = \frac{\pi}{2}, \quad \arctan\left( \frac{2q + c}{\sqrt{4d - c^2}} \right) \Big|_{q=0} = \arctan\left( \frac{c}{\sqrt{4d - c^2}} \right) = \arctan(-k_F^2) = -\frac{\pi}{2}, \tag{27}$$

$$\text{Im}\log(q^2 + cq + d) \Big|_{q \to +\infty} = 0, \qquad \text{Im}\log(q^2 + cq + d) \Big|_{q=0} = \text{Im}\log(d) = \text{Im}\log(k_F^4) = 0. \tag{28}$$

Combine all these equations, we finally obtain

$$A(\omega) = 4\pi^2 (m_1 + m_2) \mathrm{Im} \left( \frac{m_1(\omega + i\Gamma_1) + m_2(\omega + i\Gamma_2)}{\sqrt{m_1 m_2 \delta^2(k_F) - [m_1(\omega + i\Gamma_1) + m_2(\omega + i\Gamma_2)]^2}} \right). \tag{29}$$

The density of states near the Fermi energy should be only dependent on the low energy properties of the model. It is important to notice that at two dimensions, the density of states of a quadratic band is just its effective mass:

$$A_i(\omega) = -2\pi \mathrm{Im} \int_0^{+\infty} \frac{dq}{\omega - q/(2m_i) + i\eta} = 4\pi^2 m_i \theta(\omega). \tag{30}$$

Therefore, Equation (29) can be reexpressed as

$$A(\omega) = D_0 \mathrm{Im} \frac{1}{\sqrt{\delta^2 / [4(\omega + i\Gamma_A)^2] - 1}}, \tag{31}$$

where $D_0 = D_1 + D_2$, $D_i \equiv A_i(\omega = 0)$ is the density of states at the Fermi energy in the absence of scattering and hybridization,

$$\delta = \frac{2\sqrt{m_1 m_2}}{m_1 + m_2} \delta(k_F), \ \Gamma_A = \frac{m_1 \Gamma_1 + m_2 \Gamma_2}{m_1 + m_2}, \tag{32}$$

are the real band gap (as $\mathrm{Im}\sqrt{\delta^2/(4\omega^2) - 1} = 0$ for $\omega^2 < (\delta/2)^2$), and the scattering rate weighted by the effective masses respectively. This is the Eq. (5) in the main text. Note that in this way we reproduce the band gap obtained in Eq. (5).

It is also instructive to express Eq. (32) using Fermi velocities $v_i \equiv \partial \epsilon_i / \partial k(k_F) = k_F/m_i$. In this way:,

$$\delta = \frac{2\sqrt{v_1 v_2}}{v_1 + v_2} \delta(k_F), \ \Gamma_A = \frac{v_2 \Gamma_1 + v_1 \Gamma_2}{v_1 + v_2}, \tag{33}$$

Equation (33) can also be derived directly from the generic two-band Hamiltonian Eq. (1) by expanding $\epsilon_i(k) = \epsilon_i(k_F) + v_i(k - k_F)$ and $\Delta(\mathbf{k}) = \delta(k_F)$.

## V. LOW ENERGY DENSITY OF STATES UNDER MAGNETIC FIELD

### A. Landau Level Spectrum

If out of plane magnetic field is applied, the energy bands become Landau levels. With Pierels substitution, $\mathbf{k} \rightarrow \mathbf{k} - \mathbf{A}$, where $\mathbf{A}$ is the vector potential that satisfies $\nabla \times \mathbf{A} = \mathbf{B}$. Here we have already taken $e = c = 1$ for simplicity. After taking the symmetric gauge, we have the following identification of the operators:

$$k_x - ik_y = \sqrt{2B}\hat{a}, \ k_x + ik_y = \sqrt{2B}\hat{a}^\dagger, \tag{34}$$

$$k_x^2 + k_y^2 = 2B \left( \hat{a}^\dagger \hat{a} + \frac{1}{2} \right). \tag{35}$$

where $\hat{a}, \hat{a}^\dagger$ are lowering and raising operators for the quantum harmonic oscillator energy eigenstates: $\hat{a}\,|n\rangle = \sqrt{n}\,|n-1\rangle$ and $\hat{a}^\dagger\,|n\rangle = \sqrt{n+1}\,|n+1\rangle$. The noninteracting Hamiltonian then becomes

$$H_0 = \begin{pmatrix} \frac{B}{m_1}\left(\hat{a}^\dagger \hat{a} + \frac{1}{2}\right) + \frac{g}{2} + \mu_0 & \sqrt{2B}v\hat{a} \\ \sqrt{2B}v\hat{a}^\dagger & -\frac{B}{m_2}\left(\hat{a}^\dagger \hat{a} + \frac{1}{2}\right) - \frac{g}{2} + \mu_0 \end{pmatrix}. \tag{36}$$

For $n = 1$, we solve the quasiparticle Hamiltonian $H = H_0 + \Sigma$ with ansatz $(\psi_1, \psi_2) = (0, |0\rangle)$. The energy eigenvalue is

$$E_{n=0} = \mu_0 - \frac{g}{2} - i\Gamma_2 - \frac{B}{2m_2}, \tag{37}$$

which increases linearly with magnetic field, and does not contribute to the quantum oscillation.

For $n \geq 1$, we solve the quasiparticle Hamiltonian with ansatz $(\psi_1, \psi_2) = (c_1 |n-1\rangle, c_2 |n\rangle)$. The energy eigenvalues are

$$E_{n \geq 1, \pm} = \frac{1}{2}\left(\epsilon_{1,n} - \epsilon_{2,n} - i\Gamma \pm \sqrt{\left[(\epsilon_{1,n} + \epsilon_{2,n}) - i\gamma\right]^2 + v^2(8nB)}\right),$$ (38)

where

$$\epsilon_{1,n} = \frac{B}{m_1}\left(n - \frac{1}{2}\right) + \frac{g}{2} + \mu_0, \ \ \epsilon_{2,n} = \frac{B}{m_2}\left(n + \frac{1}{2}\right) + \frac{g}{2} - \mu_0.$$ (39)

For large Landau level index $n \gg 1$, Eq. (15) becomes Eq. (38) by replacing $k \to \sqrt{2nB}$. This is the semi-classical regime of the Landau level spectrum.

Apart from the Landau level index $n$, the angular momentum is another good quantum number under the symmetric gauge. This leads to a $D = B/(2\pi)$ degeneracy for each Landau level per unit area of the system.

Finally, we remark the effect of electron spins. We have been focused on $\sigma_z s_z = -1$ sector of the full Hamiltonian. For the $\sigma_z s_z = 1$ sector, the Hamiltonian is $H_0^T$. The Landau level spectrum can be similarly obtained as

$$E_{n=0} = \mu_0 + \frac{g}{2} - i\Gamma_1 + \frac{B}{2m_1},$$ (40)

and Eq. (38) with

$$\epsilon_{1,n} = \frac{B}{m_1}\left(n + \frac{1}{2}\right) + \frac{g}{2} + \mu_0, \ \ \epsilon_{2,n} = \frac{B}{m_2}\left(n - \frac{1}{2}\right) + \frac{g}{2} - \mu_0.$$ (41)

Note that in the semi-classical regime, the Landau level spectra of both $H^\uparrow$ and $H^\downarrow$ are the same. Since in the following, we will carry out our calculation in this regime (weak magnetic field limit), it is safe to ignore the difference between $H^\uparrow$ and $H^\downarrow$.

## B. Low Energy Density of States

With magnetic field, the spectral function is computed by (neglecting lowest Landau level)

$$A(\omega) = -2D\text{Im}\sum_{n=1}^{\infty}\left(\frac{1}{\omega - E_{n,+}} + \frac{1}{\omega - E_{n,-}}\right),$$ (42)

where $D = B/(2\pi)$ is the degeneracy of the Landau level. This summation can be recast into an integral using Possion summation formula: $A(\omega) = \sum_{k=0}^{\infty} A_k(\omega)$, where

$$A_k(\omega) = -4D\text{Im}\int_1^{\infty} dx \cos(2\pi kx)\left(\frac{1}{\omega - E_{x,+}} + \frac{1}{\omega - E_{x,-}}\right).$$ (43)

With the assumption of small hybridization gap $\delta(k_F) \ll k_F^2/\sqrt{m_1 m_2}$ and weak magnetic field $B \ll k_F^2$, the integrand can be simplified as

$$A_k(\omega) = \text{Im}\frac{-1}{\pi\sqrt{\left[(m_1 + m_2)\omega + i(m_1\Gamma_1 + m_2\Gamma_2)\right]^2 - m_1 m_2 \delta^2(k_F)}}\int_1^{+\infty} dx \cos\left(2\pi kx\right)\left(\frac{c_+}{x - x_+} - \frac{c_-}{x - x_-}\right),$$ (44)

$$c_\pm = (m_1 + m_2)\left[(m_1 + m_2)\omega + i(m_1\Gamma_1 + m_2\Gamma_2)\right] \pm (m_1 - m_2)\sqrt{\left[(m_1 + m_2)\omega + i(m_1\Gamma_1 + m_2\Gamma_2)\right]^2 - m_1 m_2 \delta^2(k_F)},$$ (45)

$$x_\pm = \frac{k_F^2 + (m_1 - m_2)\omega + i(m_1\Gamma_1 - m_2\Gamma_2) \pm \sqrt{\left[(m_1 + m_2)\omega + i(m_1\Gamma_1 + m_2\Gamma_2)\right]^2 - m_1 m_2 \delta^2(k_F)}}{2B}.$$ (46)

Note that by assuming weak magnetic field, we are in fact in the semi-classical regime and the difference between $H^\uparrow$ and $H^\downarrow$ can be neglected.

Since we have already assumed weak magnetic field, the real parts of $x_\pm$ are deep inside the right half of the complex plane. In this way, we can extend the integral range from $[1, +\infty)$ to $(-\infty, +\infty)$, so that the integral in Eq. (44) can be evaluated using residue theorem as long as it is not $\Gamma_1 = \Gamma_2 = 0$. Note the imaginary parts of $x_+$ and $x_-$ are always of opposite signs, the integral contour (upper or lower plane) for $x_+$ and $x_-$ are opposite. Denote $s \equiv \operatorname{sgn} \operatorname{Im} x_+ = \operatorname{sgn} m_1$. The integral is evaluated as

$$\int_{-\infty}^{+\infty} dx \cos(2\pi kx)\left(\frac{c_+}{x - x_+} - \frac{c_-}{x - x_-}\right) = s\pi i\left(c_+ e^{2\pi iksx_+} + c_- e^{-2\pi iksx_-}\right). \tag{47}$$

Particularly inside the hybridization gap, the full form of the density of states is

$$A_k(\omega = 0) = -2\cos\left(\frac{\pi kk_F^2}{B}\right)\sum_{i=\pm} M_i \exp(-\pi k\mathcal{D}_i), \tag{48}$$

with

$$M_\pm = \frac{s}{2}\left[\frac{(m_1 + m_2)(m_1\Gamma_1 + m_2\Gamma_2)}{\sqrt{(m_1\Gamma_1 + m_2\Gamma_2)^2 + m_1 m_2\delta^2(k_F)}} \pm (m_1 - m_2)\right], \tag{49}$$

$$\mathcal{D}_\pm = \frac{s}{B}\left[\sqrt{(m_1\Gamma_1 + m_2\Gamma_2)^2 + m_1 m_2\delta^2(k_F)} \pm (m_1\Gamma_1 - m_2\Gamma_2)\right], \tag{50}$$

being "effective masses" and Dingle exponents respectively. Note that the real part of $x_\pm$ determines the oscillation period and the imaginary part determines the Dingle factor. $c_\pm$ gives the "effective mass". Also note that when the magnetic field is strong $B \sim k_F^2$, one has to expand Eq. (46) to the order of $(k_F^2/B)^{-1}$, which implies a field-dependent phase shift [3–6].

At zero hybridization gap limit $\delta(k_F) = 0$, the "effective masses" and the Dingle exponents become

$$M_+ = sm_1, \; M_- = sm_2, \tag{51}$$

$$\mathcal{D}_+ = 2sm_1\Gamma_1, \; \mathcal{D}_- = 2sm_2\Gamma_2, \tag{52}$$

which are exactly the effective masses and the Dingle exponents in normal metals.

It is instructive to rewrite Eq. (49) and (50) using $\Gamma_{A,D}$ and the indirect band gap $\delta$:

$$M_\pm = \frac{s(m_1 + m_2)}{2}\left[\frac{1}{\sqrt{1 + \delta^2/(4\Gamma_A^2)}} \pm \frac{m_1 - m_2}{m_1 + m_2}\right], \tag{53}$$

$$\mathcal{D}_\pm = \frac{s(m_1 + m_2)}{B}\left(\Gamma_A\sqrt{1 + \delta^2/(4\Gamma_A^2)} \pm \Gamma_D\right), \tag{54}$$

where $\Gamma_A$ and $\delta$ are defined in Eq. (32), and $\Gamma_D \equiv (m_1\Gamma_1 + m_2\Gamma_2)/(m_1 + m_2)$. These are Eq. (9) and (10) in the main text.

To further confirm that $M_i$ is really the effective mass extracted from the temperature dependence of the oscillation amplitude, we explicitly compute the density of states at finite temperature, defined as

$$D(\omega, T) = -\int_{-\infty}^{+\infty} dE \frac{\partial n_F(E - \omega, T)}{\partial E} A(E) = \frac{1}{2T}\int_{-\infty}^{+\infty} \frac{dE}{1 + \cosh\left(\frac{E - \omega}{T}\right)} A(E), \tag{55}$$

where $A(E)$ is the spectral function evaluated in Eq. (47). Here we have taken $k_B = 1$ for simplicity. Assume the temperature is small $T \ll \delta(k_F), B/m_{1,2}$, it suffices to expand $A(E)$ around $E = 0$, i.e., expand $x_\pm$ to the leading order of $\omega$ and expand $c_\pm$ to the zeroth order of $\omega$:

$$x_\pm = x_{\pm,0} + \frac{\omega}{2B}\left[(m_1 - m_2) \pm \frac{(m_1 + m_2)(m_1\Gamma_1 + m_2\Gamma_2)}{\sqrt{(m_1\Gamma_1 + m_2\Gamma_2)^2 + m_1 m_2\delta^2(k_F)}}\right], \tag{56}$$

where $x_{\pm,0}$ is Eq. (46) at $\omega = 0$. With the helpful of the Fourier transform

$$\int_{-\infty}^{\infty} dz \frac{e^{i\lambda z}}{1 + \cosh(bz)} = \frac{2\pi\lambda}{b^2}\frac{1}{\sinh(\pi\lambda/b)}, \tag{57}$$

the integral Eq. (55) can be performed analytically. The final result is

$$D_k(\omega = 0, T) = -2\cos\left(\frac{\pi k k_F^2}{B}\right)\sum_{i=\pm}\frac{M_i\frac{2\pi^2 kT}{B/M_i}}{\sinh\left(\frac{2\pi^2 kT}{B/M_i}\right)}\exp(-\pi\mathcal{D}_i).$$ (58)

By doubling the density of states due to the contribution from $s =\uparrow$ spin sector, we get $D_1(\omega = 0, T)$ is Eq. (8) in the main text. The temperature dependence is exactly the Lifshitz-Kosevich factor for normal metals, with effective masses renormalized as Eq. (49).

---

[1] C. W. Groth, M. Wimmer, A. R. Akhmerov, J. Tworzydło, and C. W. J. Beenakker, "Theory of the Topological Anderson Insulator," Phys. Rev. Lett. **103**, 196805 (2009).

[2] Q. Y. Chen, D. F. Xu, X. H. Niu, J. Jiang, R. Peng, H. C. Xu, C. H. P. Wen, Z. F. Ding, K. Huang, L. Shu, Y. J. Zhang, H. Lee, V. N. Strocov, M. Shi, F. Bisti, T. Schmitt, Y. B. Huang, P. Dudin, X. C. Lai, S. Kirchner, H. Q. Yuan, and D. L. Feng, "Direct observation of how the heavy-fermion state develops in CeCoIn$_5$," Phys. Rev. B **96**, 045107 (2017).

[3] Dong-Xia Qu, Y. S. Hor, Jun Xiong, R. J. Cava, and N. P. Ong, "Quantum Oscillations and Hall Anomaly of Surface States in the Topological Insulator Bi$_2$Te$_3$," Science **329**, 821–824 (2010).

[4] Zhi Ren, A. A. Taskin, Satoshi Sasaki, Kouji Segawa, and Yoichi Ando, "Large bulk resistivity and surface quantum oscillations in the topological insulator Bi$_2$Te$_2$Se," Phys. Rev. B **82**, 241306 (2010).

[5] Anthony R. Wright and Ross H. McKenzie, "Quantum oscillations and Berry's phase in topological insulator surface states with broken particle-hole symmetry," Phys. Rev. B **87**, 085411 (2013).

[6] E. Tisserond, J. N. Fuchs, M. O. Goerbig, P. Auban-Senzier, C. Mézière, P. Batail, Y. Kawasugi, M. Suda, H. M. Yamamoto, R. Kato, N. Tajima, and M. Monteverde, "Aperiodic quantum oscillations of particle-hole asymmetric Dirac cones," EPL (Europhysics Letters) **119**, 67001 (2017).


\end{document}